\documentclass[journal]{IEEEtran}

\usepackage[utf8]{inputenc}
\usepackage{color}
\usepackage{xcolor}
\usepackage{array}
\usepackage{verbatim}
\usepackage{float}
\usepackage{amsmath}
\usepackage{amsthm}
\usepackage{amssymb}
\usepackage{graphicx}
\usepackage{longtable}
\usepackage{multirow}
\usepackage{booktabs}
\usepackage[unicode=true,
bookmarks=false,
breaklinks=false,pdfborder={0 0 1},colorlinks=false]
{hyperref}
\hypersetup{
	colorlinks,bookmarksopen,bookmarksnumbered,citecolor=blue,urlcolor=blue}
\usepackage{cite}

\usepackage{lipsum}
\usepackage{mathtools}
\usepackage{cuted}
\providecommand{\tabularnewline}{\\}
\usepackage{algorithmic}
\usepackage{longtable}

\floatstyle{ruled}
\newfloat{algorithm}{tbp}{loa}
\providecommand{\algorithmname}{Algorithm}
\floatname{algorithm}{\protect\algorithmname}

\makeatletter
\let\oldforeign@language\foreign@language
\DeclareRobustCommand{\foreign@language}[1]{%
	\lowercase{\oldforeign@language{#1}}}

\let\oldforeign@language\foreign@language
\DeclareRobustCommand{\foreign@language}[1]{%
	\lowercase{\oldforeign@language{#1}}}

\ifCLASSINFOpdf
\else
\fi

\@ifundefined{showcaptionsetup}{}{%
	\PassOptionsToPackage{caption=false}{subfig}}
\usepackage{subfig}

\usepackage{balance}

\ifCLASSINFOpdf
\else
\fi

\ifCLASSINFOpdf
\else
\fi

\def\ps@IEEEtitlepagestyle{%
	\def\@oddhead{\parbox[t][\height][t]{\textwidth}{\centering \scriptsize
			Personal use of this material is permitted. Permission from the author(s) and/or copyright holder(s), must be obtained for all other uses. Please contact us and provide details if you believe this document breaches copyrights.\\
			\noindent\makebox[\linewidth]{}
		}\hfil\hbox{}}%
	\def\@evenhead{\scriptsize\thepage \hfil \leftmark\mbox{}}%
	\def\@oddfoot{\parbox[t][\height][l]{\textwidth}{
			\vspace{-20pt}{\rule{\textwidth}{0.4pt}}\\ \footnotesize{\bf{\footnotesize\textcolor{red}{A. Yu, J. Kolotylo, H. A. Hashim, and A. E.E. Eltoukhy, "Electronic Warfare Cyberattacks, Countermeasures and Modern Defensive Strategies of UAV Avionics: A Survey," IEEE Access, pp. 1-20, 2025.}}}\\\\
			\noindent\makebox[\linewidth]
		}\hfil\hbox{}}%
	\def\@evenfoot{\MYfooter}}

\makeatother
\begin{document}
	\bstctlcite{IEEEexample:BSTcontrol}

\title{Electronic Warfare Cyberattacks, Countermeasures and Modern Defensive Strategies of UAV Avionics: A Survey}

\author{Aaron Yu, Iuliia Kolotylo, Hashim A. Hashim, and A. E.E. Eltoukhy
	\thanks{This work was supported in part by the National Sciences and Engineering Research Council of Canada (NSERC), under the grants RGPIN-2022-04937.}
	\thanks{A. Yu, I. Kolotylo, and H. A. Hashim are with the Department of Mechanical and Aerospace Engineering, Carleton University, Ottawa, ON, K1S-5B6, Canada. A. E.E. Eltoukhy is with Management Science and Engineering Department, Khalifa University, Abu Dhabi, United Arab Emirates. Corresponding author contact (e-mail: hhashim@carleton.ca).}
}



\maketitle
\begin{abstract}
Unmanned Aerial Vehicles (UAVs) play a pivotal role in modern autonomous air mobility, and the reliability of UAV avionics systems is critical to ensuring mission success, sustainability practices, and public safety. The success of UAV missions depends on effectively mitigating various aspects of electronic warfare, including non-destructive and destructive cyberattacks, transponder vulnerabilities, and jamming threats, while rigorously implementing countermeasures and defensive aids. This paper provides a comprehensive review of UAV cyberattacks, countermeasures, and defensive strategies. It explores UAV-to-UAV coordination attacks and their associated features, such as dispatch system attacks, Automatic Dependent Surveillance-Broadcast (ADS-B) attacks, Traffic Alert and Collision Avoidance System (TCAS)-induced collisions, and TCAS attacks. Additionally, the paper examines UAV-to-command center coordination attacks, as well as UAV functionality attacks. The review also covers various countermeasures and defensive aids designed for UAVs. Lastly, a comparison of common cyberattacks and countermeasure approaches is conducted, along with a discussion of future trends in the field.
\end{abstract}

\begin{IEEEkeywords}
Electronic warfare, UAVs, Avionics Systems, cyberattacks, coordination
attacks, functionality attacks, countermeasure, defensive-aids.
\end{IEEEkeywords}

\begin{table}[h]
	\caption{\label{tab:Nomenclature}Nomenclature.}
	
	\centering{}%
	\begin{tabular}{>{\raggedright}p{1.3cm}>{\raggedright}p{6.5cm}}
		\toprule 
		\addlinespace[0.1cm]
		ADS-B & Automatic Dependent Surveillance - Broadcast\tabularnewline
		\addlinespace[0.1cm]
		ATC & Air Traffic Control\tabularnewline
		\addlinespace[0.1cm]
		COMINT & Communications Intelligence\tabularnewline
		\addlinespace[0.1cm]
		DoS & Denial-of-Service\tabularnewline
		\addlinespace[0.1cm]
		DSSS & Direct Sequence Spread Spectrum\tabularnewline
		\addlinespace[0.1cm]
		EA & Electronic Attack\tabularnewline
		\addlinespace[0.1cm]
		EASA & European Union Aviation Safety Agency\tabularnewline
		\addlinespace[0.1cm]
		ELINT & Electronic Intelligence\tabularnewline
		\addlinespace[0.1cm]
		EMS & Electromagnetic Spectrum\tabularnewline
		\addlinespace[0.1cm]
		EP & Electronic Protection\tabularnewline
		\addlinespace[0.1cm]
		ES & Electronic Warfare Support\tabularnewline
		\addlinespace[0.1cm]
		FAA & Federal Aviation Administration\tabularnewline
		\addlinespace[0.1cm]
		FH & Frequency Hopping\tabularnewline
		\addlinespace[0.1cm]
		FHSS & Frequency Hopping Spread Spectrum\tabularnewline
		\addlinespace[0.1cm]
		FL & Federated Learning\tabularnewline
		\addlinespace[0.1cm]
		GNSS & Global Navigation Satellite Systems\tabularnewline
		\addlinespace[0.1cm]
		GPS & Global Positioning System\tabularnewline
		\addlinespace[0.1cm]
		IFF & Identify Friend or Foe\tabularnewline
		\addlinespace[0.1cm]
		IMU & Inertial Measurement Unit\tabularnewline
		\addlinespace[0.1cm]
		ICAO & International Civil Aviation Organization\tabularnewline
		\addlinespace[0.1cm]
		LiDAR & Light Detection and Ranging\tabularnewline
		\addlinespace[0.1cm]
		LTE & Long-Term Evolution\tabularnewline
		\addlinespace[0.1cm]
		MAC & Media Access Control\tabularnewline
		\addlinespace[0.1cm]
		NMAC & Near Midair Collision\tabularnewline
		\addlinespace[0.1cm]
		QKD & Quantum Key Distribution\tabularnewline
		\addlinespace[0.1cm]
		RID & Remote Identification\tabularnewline
		\addlinespace[0.1cm]
		SATCOM & Satellite Communications\tabularnewline
		\addlinespace[0.1cm]
		SDR & Software-defined Radio\tabularnewline
		\addlinespace[0.1cm]
		SIGINT & Signal Intelligence\tabularnewline
		\addlinespace[0.1cm]
		TCAS & Traffic Alert and Collision Avoidance System\tabularnewline
		\addlinespace[0.1cm]
		TCP/UDP & Control Protocol/User Datagram Protocol\tabularnewline
		\addlinespace[0.1cm]
		UWB & Ultra-wideband\tabularnewline
		\addlinespace[0.1cm]
		UAVs & Unmanned Aerial vehicles\tabularnewline
		\bottomrule
	\end{tabular}
\end{table}

\section{Introduction}

\subsection{Motivation}

Unmanned Aerial Vehicles (UAV) have become increasingly important
in both civilian and military applications over the past several decades.
Within the civil sector, UAVs have seen significant growth, finding
use in news broadcasting, agriculture, construction, and other business
ventures \cite{hashim2023exponentially,ref02,ramachandran2021review,hashim2023uwbITS}.
Military UAVs have also been used extensively in surveillance and
object targeting \cite{hashim2023exponentially,ref01}. These prospects
are made possible by the major trends in UAV development, with current
research targeting increasing levels of autonomy through advancements
in on-board avionics systems \cite{ref03,wanner2024uav}. The general
UAV avionics composition is similar to those found on manned aircrafts
and can be broken into several subsystems: propulsion, electrical,
sensors, communications, flight navigation, and control \cite{ref04}.
The avionics system interacts with these subsystems for different
purposes, such as controlling and maintaining flight altitude with
the propulsion system, drawing power from the batteries of the electrical
system, or collecting data from the sensor system. The integration
and control of these processes is a complex task that often relies
on one or more flight computers \cite{hashim2023exponentially,ref04,wanner2024uav}.
Communications between UAVs (air-to-air) and to ground stations (air-to-ground)
is a critical component of UAV networks, as it provides the proper
means of identification and communication for drones to accomplish
their individual tasks. Some of the commonly used notation are listed
in Table \ref{tab:Nomenclature}.

\begin{figure*}[h]
	\centering{}\includegraphics[scale=0.54]{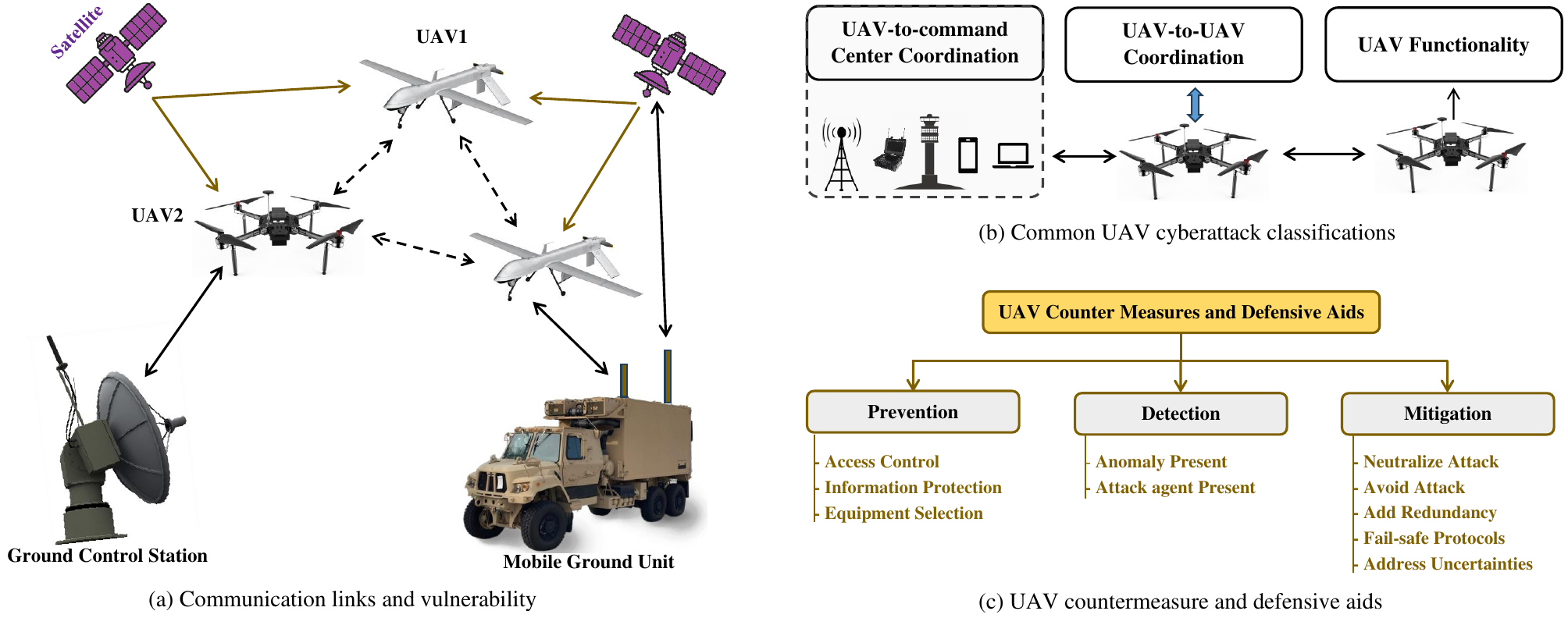}\caption{\label{fig:Common-UAV-cyberattack}Conceptual illustration of UAV
		electronic warfare: (a) Communication links, (b) common UAV cyberattack
		classifications, and (c) countermeasures and defensive aids.}
\end{figure*}

\subsection{Electronic Warfare and UAVs}

Electronic warfare encompasses the science and art of maintaining
control over the Electromagnetic Spectrum (EMS) for friendly use while
denying its use to adversaries. In the context of UAVs, this primarily
involves military tactics, such as information gathering and disruption
of operations by exploiting communication links within UAV networks,
as illustrated in Figure \ref{fig:Common-UAV-cyberattack}.(a). This
area of warfare is an expanding field of research, with a growing
focus on developing strategies and techniques that can both protect
friendly UAVs from hostile actors and disrupt the operations of adversarial
UAVs. For civilian UAVs, this type of warfare is more commonly referred
to in broader scopes in relation to cybersecurity and cyberattacks
\cite{hartmann2013vulnerability,mitchell2013adaptive}. Along with
UAVs, electronic warfare began development during World War II and
picked up significantly during following conflicts. Jamming techniques
were first explored in 1902 by the Royal Navy with the introduction
of the radio. They used for the first time in combat by a Russian
operator during the 1905 against Russo-Japanese war by overlapping
signals meant for the correction of Japanese naval gunfire \cite{ref18}.
Once RADAR was established in England, 1935, further research into
jamming began along with the official introduction of electronic warfare.
This led to the development of RADAR reflectors used in 1943 to confuse
German RADAR operators and win British victories. They were used again
in 1945 to completely hide U.S. aircraft from Japanese searchlight
RADARs \cite{riabukha2020radar}. However, post-war electronic warfare
development halted for a half-decade as engineers returned to civilian
life and most military equipment were sold off. No U.S. study on lessons-learned
for electronic warfare was produced, and thus most successes and developments
were never fully documented. The following conflicts involving the
U.S. were limited in their electronic warfare capabilities due to
this gap in knowledge \cite{ref18}. The extent of electronic warfare
in the Korean War was the U.S. B-12s jamming Chinese RADAR systems.
Nevertheless, advancements in UAV technology and electronic warfare
have continued into the twenty-first century, resulting in modernized
capabilities.

\subsection{Cyberattacks Related to UAVs}

Due to the increasing use of UAVs for commercial applications, cyberattacks
are a concern for both military and civilian class drones \cite{ref47}.
Various classifications of UAV related cyberattacks exist, with some
examples being destructive and non-destructive attacks \cite{ref46},
ground-based and air-based attacks \cite{ref49}, and threat categorized
attacks ranging from physical, sensor, and communication, to supply
chain threats \cite{ref50}. The classifications can then further
be characterized by their military impact as opposed to civilian impacts.
Because of the rapidly growing development of UAVs in the public sector,
most papers tend to focus on the civilian side of UAV cyberattacks
and their repercussions. Drone cyberattacks can be split into three
classes: UAV-to-UAV coordination attacks, UAV-to-command center coordination
attacks, and UAV functionality attacks. In civil applications, drones
are often deployed in groups to carry out tasks. Even singular UAVs
typically share the air space with other aerial vehicles, which can
include manned aircraft alongside other drones on separate missions.
As a result, UAV-to-UAV coordination attacks involve interfering in
the messaging systems amongst UAVs. In serious cases, such attacks
can lead to midair collisions among other issues \cite{ref51}. Commonly
termed attacks that belong to this category include dispatch system
attacks, Automatic Dependent Surveillance - Broadcast (ADS-B) attacks,
and Traffic Alert and Collision Avoidance System (TCAS) induced collisions.
In a UAV network, the command center is responsible for keeping track
of and maintaining each UAV’s operational status such as payload control,
mission planning, and air vehicle control \cite{hashim2023uwbITS}.
This information is transmitted to the UAV through wireless communication
links that, because of historical development in transponder technology,
is not inherently secure. Cyberattacks of this nature passively or
actively disrupt the communication link between the UAV and command
center. Passive attacks, such as eavesdropping, focus on obtaining
information from the communication link for malicious purposes, while
active attacks such as jamming actively interfere with the UAV’s ability
to communicate to the command center. Unlike the previous two categorizations,
cyberattacks targeting the UAV’s functionality precisely alter the
way the UAV behaves as opposed to interfering with its communication
channels \cite{ref51,humphreys2013detection,jafarnia2012gps}, although
the methods of access can be similar. This is dependent on the particular
UAV’s hardware configuration; however common examples include exploitation
of recorded video attacks and Global Positioning System (GPS) spoofing.
In such attacks, fake signals are generated by malicious actors which
are then sent to the UAV. The UAV believes that the fake information
is genuine, altering the way it executes its mission depending on
the type of attack being performed. In many cases, this attack tricks
the UAV into believing its spoofed location is legitimate, allowing
attackers to hijack the UAV’s flight path and steal it. Figure \ref{fig:Common-UAV-cyberattack}.(b)
shows categorization of various cyberattacks on UAVs exposed by UAV
systems in public as well as military drones.

\subsection{Countermeasures and Defensive-Aids}

Electronic countermeasure systems employ specific tactics, techniques,
and technologies to interfere with, deceive, disrupt, degrade, or
neutralize an adversary's electronic and radar systems of the enemy
\cite{jo2021survey,ref73,ref74}. Electronic counter-countermeasure
systems, as part of defensive aid measures, are designed to counteract
electronic countermeasures by restoring radar functionality and mitigating
their effects \cite{zohuri2020electronic}. Additionally, electronic
counter-countermeasure systems systems encompass strategies to protect
friendly systems from electronic threats, such as encryption and signal
shielding. Electronic countermeasure and electronic counter-countermeasure
are known as cat and mouse game. The most common threats faced by
aircraft are small arms fire, radar guided anti-aircraft missiles,
shoulder-launched surface-to-air missiles (SAM), and SAM mounted to
ground sites, vehicles or ships \cite{sonawane2011tactical}. On the
other side, the defensive-aid subsystem related to aircrafts would
include radar warning receiver, missile warning receiver, laser warning
receiver, countermeasure dispenser (chaff or flares), and towed decoy
\cite{neri2006introduction}. With regard to UAVs, countermeasures
and defensive-aids are methods applied to protect the data handled
by UAVs, more specifically, confidentiality, integrity, and authenticity
\cite{ref71,ref73,ref74,ref79,ref80,ref84,li2024secure,li2025adaptive}.
Also, countermeasures and defensive-aids for UAVs are used to ensure
the service availability for civilian and military applications \cite{naik2020dilemma}.
In view of UAV cyberattacks, a wide spectrum of countermeasures are
created to take advantage of vulnerabilities found on hardware, software,
and network layers \cite{hashim2023exponentially,hashim2023observer}.
The same countermeasure may work against several types of attacks
covered under three broad categories: prevention, detection, and mitigation
(as shown in Figure \ref{fig:Common-UAV-cyberattack}.(c)).

\subsection{Other Impacts of Electronic Warfare }

Electronic warfare often involves the use of high-powered electronic
signals, which can contribute to electromagnetic pollution, potentially
harming the environment and living organisms, and negatively impacting
sustainability efforts \cite{naik2020dilemma}. High levels of electromagnetic
radiation may interfere with the navigation systems of wildlife, leading
to behavioral disruptions. In terms of energy consumption, electronic
warfare systems, such as jammers, radars, and signal intelligence
receivers, demand substantial energy to operate. These systems frequently
rely on portable generators or ground/aerial vehicles, resulting in
increased fuel consumption, greenhouse gas emissions, and subsequent
environmental degradation. Additionally, electronic warfare can disrupt
civilian communication and navigation systems, including emergency
services and GPS-based systems \cite{hashim2023uwbITS}. This disruption
can lead to an increased risk of accidents in sensitive areas and
a reduction in the effectiveness of environmental monitoring and response
systems.

\subsection{Focus and Structure}

\paragraph*{Scope}This survey article aims to provide a comprehensive
overview of UAV electronic warfare, encompassing cyberattacks, countermeasures,
and defensive aids. It offers a detailed characterization of various
attack taxonomies and systematically reviews UAV-to-UAV coordination
attacks, classifying them into distinct categories such as dispatch
system attacks (including message elimination, message spoofing, and
message fabrication), ADS-B attacks, TCAS-induced collisions, and
TCAS attacks. Additionally, UAV-to-command center coordination attacks
are analyzed, covering threats such as eavesdropping, man-in-the-middle
attacks, jamming, and Wi-Fi-based intrusions. UAV functionality attacks
are also explored, addressing risks such as recorded video exploitation,
Denial-of-Service (DoS) attacks, and GPS spoofing. Beyond categorizing
cyber threats, this article examines UAV electronic warfare in both
civil and military contexts, discussing countermeasures and defensive
aids related to prevention techniques, information security, communication
traffic management, mitigation strategies, and the avoidance of wireless
communication vulnerabilities. Furthermore, a comparative analysis
of common cyberattacks and countermeasure approaches is presented,
along with an exploration of emerging trends in the field. Unlike
prior works such as \cite{jo2021survey,mpitziopoulos2009survey,pirayesh2022jamming},
which focus on network attacks and the corresponding countermeasures
for communication signals, this study is dedicated specifically to
UAV electronic warfare, including cyberattacks, countermeasures, and
modern defensive-aid techniques. The studies in \cite{ref06,ref56,ref58}
primarily address identification threats but do not extensively examine
UAV-to-UAV coordination attacks, UAV-to-command center coordination
attacks, or functionality-based attacks. Similarly, \cite{ref66}
investigates various types of UAV communication attacks but lacks
an in-depth discussion of countermeasures and defensive-aid strategies.
Furthermore, the studies in \cite{ref70,ref88} emphasize communication
network security, software security analysis, and intelligent security,
with a primary focus on functionality attacks. However, they do not
comprehensively explore the role of transponders in security threats,
the distinctions between security challenges in manned aircraft and
UAVs, or UAV-to-UAV coordination attacks. The works in \cite{ref89,shafique2021survey}
are primarily structured around cryptographic methods for securing
communications and developing security protocols. Therefore, this
study provides a more extensive and holistic discussion of UAV-to-UAV
coordination attacks including those involving transponders UAV-to-command
center coordination attacks, UAV functionality attacks, and their
associated countermeasures and defensive aids. Additionally, it presents
an analysis of current challenges and outlines future research directions
in the domain of UAV electronic warfare.

\paragraph*{Structure}The rest of the paper is composed of eight
sections. Section \ref{sec:UAV-to-UAV} presents UAV-to-UAV coordination
attacks. Section \ref{sec:UAV-to-Command} discusses UAV-to-command
center coordination attacks. Section \ref{sec:UAV-Functionality-Attacks}
summarizes UAV functionality attacks. Section \ref{sec:UAV-Military}
discusses UAV electronic warfare for military purposes. Section \ref{sec:Countermeasures-and-Defensive}
presents countermeasures and popular defensive-aids. Section \ref{sec:Comparison-of-UAV}
provides comparison of UAV cyberattacks and countermeasures strategies.
Section \ref{sec:Future-Trends} presents future trends. Finally,
Section \ref{sec:Conclusion} concludes the work.

\section{UAV-to-UAV Coordination Attacks\label{sec:UAV-to-UAV}}

\subsection{Transponders Overview}

The purpose of Identify Friend or Foe (IFF) is to identify friendly
and nonfriendly aircraft as the number of aircraft in airspace increases.
Generally, this consists of an airborne receiver that listened for
primary radar transmissions and would reply, at the same fSrequency,
with a message specific to that aircraft \cite{ref11}. TCAS is an
avionics system that relies on transponders to perform air-to-air
interrogation and warns the pilot about dangerous encounters with
nearby aircraft \cite{ref27}. This is a cooperative system, meaning
that for it to function all aircraft should be equipped with a mode
S transponder to perform surveillance (Mode S surveillance). Mode
C surveillance can also be performed by TCAS to interrogate aircraft
with Mode A/C transponders \cite{ref28}. TCAS sends out pulses and
receives responses from neighboring aircraft transponders which contain
range, altitude, and azimuth (bearing). When the system detects other
aircraft that are at safe range and altitude away the pulses are sent
once a minute to perform passive surveillance. Passive surveillance
continues when the spacecraft is at either close range or altitude
at higher rate of once per 10 seconds. Once the aircraft enters potential
collision zone (i.e. it is close in both altitude and range) TCAS
engages in active surveillance at rate of 1 pulse per second \cite{ref28}.
ADS-B is another transponder-based avionics system that aids aircraft
surveillance. Unlike IFF and TCAS it does not rely on interrogations
from either Air Traffic Control (ATC) or other aircraft as it automatically
broadcasts flight state parameters to everyone involved in the network
\cite{ref34}. Similar to TCAS, ADS-B is a cooperative sensing system,
meaning that in order to receive ADS-B broadcast all players must
have ADS-B system on board. ADS-B uses Mode S transponder and broadcasts
identification, position, and velocity at frequency of 1 Hz. Position
and velocity are determined from GNSS signals in conjunction with
data from sensor systems (Inertial Measurement Unit (IMU), barometer,
etc.).

Since transponders converse by broadcasting information encoded in
electromagnetic waves, it is not surprising to believe that the electromagnetic
environment surrounding drones can become densely proliferated by
noise leading to communication issues. However, with over 400,000
expected drones on daily basis for use in commercial and government
missions in European airspace by 2050 \cite{ref21}, the matter of
bandwidth limitations between similarly broadcasting UAVs becomes
an important topic of concern \cite{ref20}. Evidently, the cumulation
of messages from increasing numbers of sources results in an electromagnetic
environment indiscernible from noise. For instance, ADS-B congestion
of the 1090 MHz band caused the Federal Aviation Administration (FAA)
to prohibit the use of ADS-B on UAVs \cite{ref20}, which in turn
has prompted research into alternative ADS-B “like” technologies to
accommodate the requirements of larger drone networks \cite{ref20,ref23}.
European Union Aviation Safety Agency (EASA) shares the same concern
with EMS usage and safety \cite{wanner2024uav}. When network congestion
occurs, transponders become more likely to fail their tasks since
interference limits their ability to decode interrogations from other
UAVs (air-to-air) or ground stations (air-to-ground).

\begin{figure}[h]
	\begin{centering}
		\includegraphics[scale=0.4]{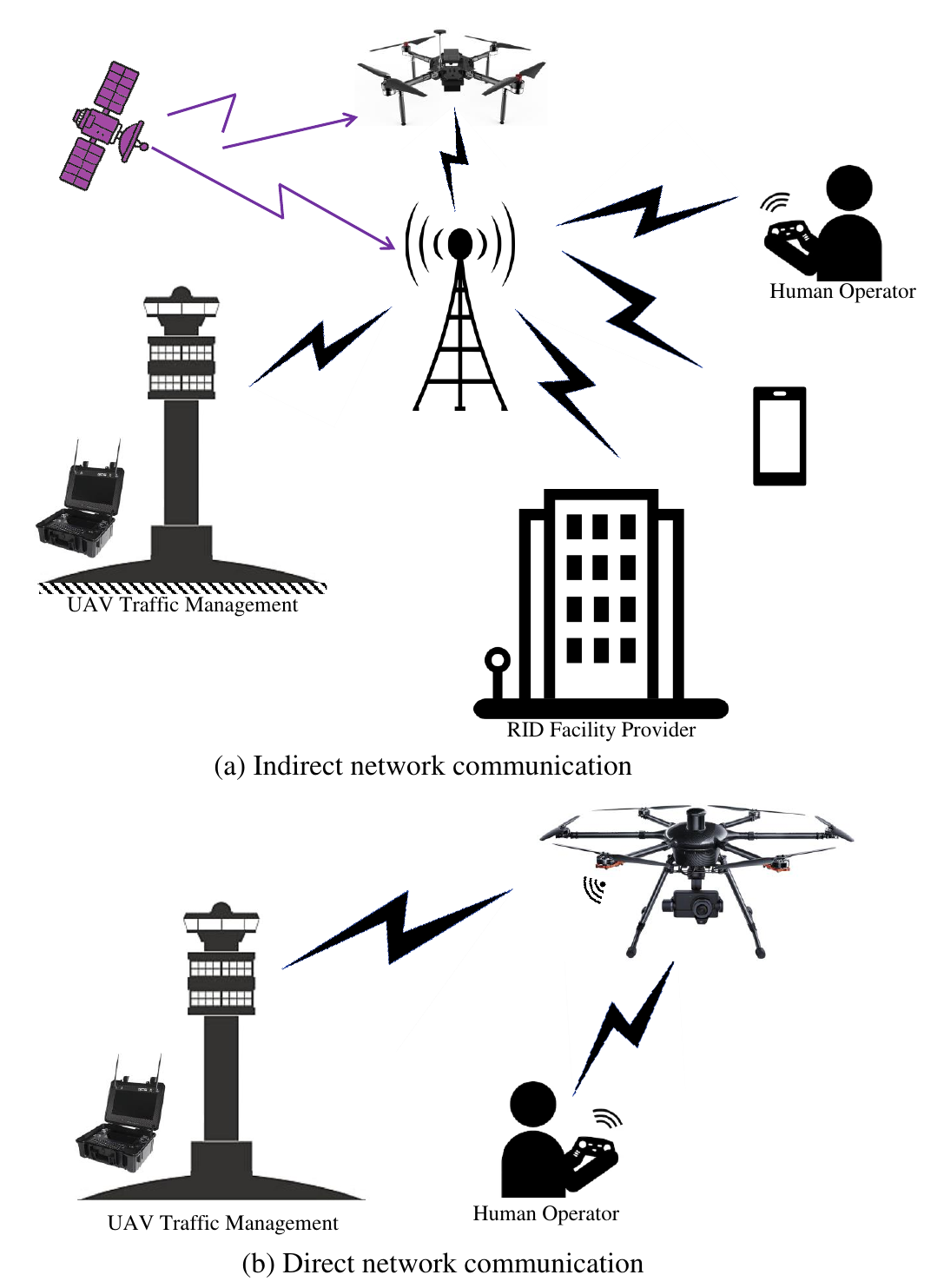}
		\par\end{centering}
	\caption{\label{fig:RID-Technologies}RID Communication Technologies.}
\end{figure}

While remote identification (RID) is a general term, in the context
of drones, it refers to the larger group of new methods of identification
for UAVs specifically that has not been historically associated with
aviation. While ADS-B, TCAS and IFF have been tried on aircraft, alternative
methods may prove more beneficial for UAVs. There are two categories
of remote identification \cite{ref06}. First are technologies that
continuously broadcast the location of the drone to local receivers.
This is done by having a GPS receiver on the drone, and a transmitter
on the drone that sends out over Wi-Fi, Bluetooth or LoRa frequencies.
The second category is network communication, which uses a cellular
network that the operator must subscribe the drone to as one would
for a cellular phone.  The broadcasting method is mainly intended
for short range and typically only includes communication from the
drone to only the operator. The appeal of these broadcasting methods
is that the relevant technology does not require a licence to purchase
or operate. It would be up to the operator’s receiving device to then
transmit the information to other parties, such as ATC or other aircraft/UAVs.
Fig. \ref{fig:RID-Technologies} presents RID communication technologies.
Using Wi-Fi as the connection can work up to 2 km if using neighbor
awareness networking and consumes 100 mW from the drone power system.
However, this has yet to be implemented on a drone. The DJI drone
company has attempted to use lower range Wi-Fi to track many drones
in an area by transmitting and receiving directly from the drone to
mobile phones. This has worked up to about 1 km. For ATC tracking
of drones, UAV-to-UAV, or UAV to aircraft identification, that information
would have to be recommunicated by the operator’s mobile device. Another
method, Bluetooth, has low bandwidth that can reach up to 200 m if
using Bluetooth 4.x or 1 km if using Bluetooth 5.x, consuming a maximum
of 10 mW. Unfily company’s Broadcast Location \& Identification Platform
(BLIP) uses Bluetooth to broadcast drone details for up to 200 m and
transfer that data through operator device’s Long-Term Evolution (LTE)
network connection. LTE is associated to fourth-generation (4G). LoRa
can reach a longer distance with a higher bit rate than Wi-Fi or Bluetooth
\cite{ref06,ref44}. Table \ref{tab:Transponder_Comp} goes into detail
about how each of the transponders compare for implementation on drones
\cite{ref11,ref27,ref28,ref34,ref06,ref44}.

\begin{table*}[h]
	\begin{centering}
		\caption{\label{tab:Transponder_Comp}Transponder comparisons in the context
			of implementation on drones.}
		\par\end{centering}
	\centering{}%
	\begin{tabular}{>{\raggedright}p{2.2cm}>{\raggedright}p{1.1cm}>{\raggedright}p{1cm}>{\raggedright}p{1.2cm}>{\raggedright}p{4.5cm}>{\raggedright}p{5.3cm}}
		\toprule 
		\addlinespace
		Transponder & Weight & Power & Access & Advantages & Disadvantages to ATC\tabularnewline\addlinespace
		\midrule
		\midrule 
		\addlinespace
		IFF/SSR & Heavy & High & Low & Has been attempted & Requires radio license and yet to be implemented\tabularnewline\addlinespace
		\midrule 
		\addlinespace
		TCAS & Heavy & High & Medium & Has been attempted & Only implemented within research context\tabularnewline\addlinespace
		\midrule 
		\addlinespace
		ADS-B & Heavy & High & Medium & Implemented in academic research & Only implemented within research context\tabularnewline\addlinespace
		\midrule 
		\addlinespace
		RID: Wi-Fi & Light & Low & High & Easy to obtain and use, no license & Short range, 2km, cannot fly drone very far\tabularnewline\addlinespace
		\midrule 
		\addlinespace
		RID: Bluetooth & Light & Low & High & Easy to obtain and use, no license & Very short range, 200m\tabularnewline\addlinespace
		\midrule 
		\addlinespace
		RID: LoRa & Light & Low & Medium & Prioritizes operator control & Requires radio license and interference probability\tabularnewline\addlinespace
		\midrule 
		\addlinespace
		RID: Cellular Network & Light & Low & Medium & More affordable and easier to implement than ADS-B & Infrastructure not ready for global coverage\tabularnewline\addlinespace
		\bottomrule
	\end{tabular}
\end{table*}

\subsection{Dispatch System Attacks }

A dispatch system refers to the hardware and software protocols and
algorithms devoted to ensuring that a group of drones with a common
objective can autonomously perform their tasks without impeding one
another \cite{ref52}. Consequently, any attack on the dispatch system
can possibly alter the behaviour of all the drones on the network.
The most common way to perform this attack is by injecting malware
into the system \cite{ref51}. The malware typically includes malicious
firmware at the software level, and trojans at the hardware level
\cite{ref54}, which are injected to produce vulnerabilities within
the UAV’s flight controller as well as the ground control station.
Software level malware can affect the UAV in countless ways that are
not limited to dispatch system attacks. An example that falls under
this category of attack is the Maldrone virus \cite{ref55}. Maldrone
was developed as the first ever backdoor malware written for AR drone
ARM Linux systems. It allows a remote hijacker to remotely switch
the drone’s software to obey their backdoor controller, providing
remote manipulation and access to the drone \cite{ref55}. Evidently,
such viruses can be devastating for a network of drones if even one
such drone becomes infected by a malicious actor. Hardware trojans,
on the other hand, include hardware level modifications to the circuit
of the flight controller. Due to the complex underlying systems that
comprise the flight controller, trojans are most commonly introduced
due to non-trusted, imitation hardware being used at some point in
the supply chain. These security breaches compromise the functions
of the circuit itself, leading to premature failure of components
and untimely destruction of the drone \cite{ref54}. In more severe
cases, the trojan can introduce backdoors that leak information to
the attackers or allow total takeover of the drone itself. An example
of such a trojan was a keylogging virus installed in a ground control
unit of the U.S. Air Force, leading to a backdoor access which malicious
actors used to track the keystrokes made in controlling the U.S. drone
fleet over Iraq and Afghanistan \cite{ref56}.

\subsection{ADS-B Attack}

ADS-B transponders are used in drones to provide autonomous collision
avoidance capabilities, which becomes more important with increasing
numbers of drones in the same air space. Attacks on the ADS-B system
involve exploiting the fact that transmitted messages are sent in
plain text format \cite{ref51}, and do not inherently provide any
authentication methods to stop message tampering \cite{ref57,chakraborty2024blocktoll}.
These attacks can further be characterized into the following three
categories: Message elimination, message infusion (spoofing), and
message fabrication.

\paragraph*{Message Elimination}It involves using external transmitters
to project constructive or destructive interference into the ADS-B
signal. When constructive interference is used, the attack induces
bit errors into the ADS-B message, causing the receiving drones to
disregard the message once it detects the manipulation and thus diminishing
awareness of the transmitter drone. When destructive interference
is used, the attacking signal is an inverse of the original ADS-B
signal, leading to complete or partial destruction of the message
\cite{ref54}.

\begin{figure}[h]
	\centering{}\includegraphics[scale=0.42]{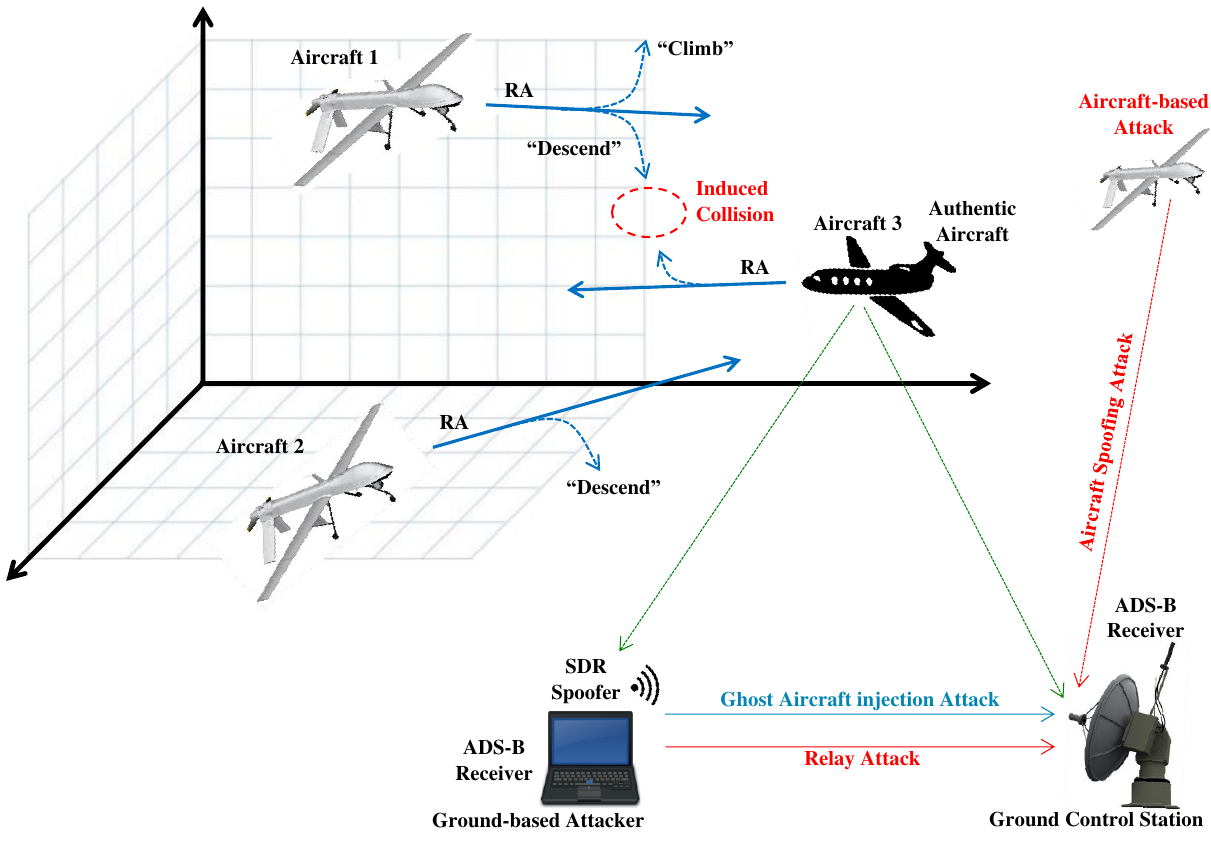}\caption{\label{fig:Spoofing}Illustration of TCAS induced collision example
		and ADS-B spoofing attacks through a ground-based attack, which replays
		a recorded ADS-B signal, and an aircraft-based attack, which spoofs
		the aircraft’s ICAO address.}
\end{figure}

\paragraph*{Message Infusion (Spoofing)}The approach of message infusion
involves injecting malicious messages into the airspace, causing ADS-B
receivers to perceive the appearance of an illegitimate aircraft.
This is possible since ADS-B does not use any authentication methods
in its messages, and so infusion can be performed by commercially
available devices. The broadcasted false messages can either target
the UAVs themselves (Aircraft Target Ghost Injection) or the ground
command center (Command Center Ghost Injection). In both cases, the
target ADS-B receiver sees a fake aircraft in the air space as the
attacker anonymously manipulates the air traffic \cite{ref52}. ADS-B
spoofing is dangerous as it can allow enemies to masquerade as potential
allies in the view of the ground station. Figure \ref{fig:Spoofing}
shows how a spoofing situation might look, where the presence of ground-based
and aircraft-based attackers infiltrate the airspace of authentic
aircraft, causing the ground station to perceive 3 different and friendly
aircraft.

\paragraph*{Message Fabrication}Message fabrication involves manipulation
of ADS-B signals to provide false information which ADS-B receivers
than interpret. Unlike message infusion, message fabrication manipulates
the messages sent by legitimate UAVs. The level of tampering can vary
based on the intention. Overshadowing is a method where the attacker
broadcasts a very high-powered ADS-B signal to substitute parts of
the ADS-B message, or entire message in the worst case. Similarly,
bit flipping involves flipping certain bits to partially manipulate
the signal. In either case, information is removed from the original
message and malicious data is inserted in its place \cite{ref51}.
These ADS-B attacks can be accomplished using commercial off-the-shelf
components as well as freely licensed software \cite{ref59}. The
most popular example of this is with Software-defined Radio (SDR)
devices, which can be programmed to transmit radio signals of different
frequencies, including those of ADS-B signals. By following the general
packet structure of an ADS-B message, Figure \ref{fig:Spoofing},
any SDR can effectively mask themselves as an ADS-B transmitter.

\subsection{TCAS Induced Collision }

Similar to ADS-B, TCAS is used to coordinate maneuvers between UAVs
in a local air space for the purpose of avoiding collisions. However,
the resolution advisories produced by TCAS do not predict long term
effects. This leads to the problem of TCAS induced collision, wherein
the resolution advisories provided by TCAS result in a collision.
Figure \ref{fig:Spoofing} shows an example of such a situation \cite{ref51}.
Aircraft pairs 1 and 2, as well as 3 and 4, are originally on collision
paths but are provided resolution advisories from TCAS. Since the
TCAS advisories were made independently, this can lead to a situation
where aircraft 4 climbs while aircraft 1 descends, leading to an induced
collision with less time to provide corrections as a consequence of
TCAS decisions. This can be extended to the case of UAVs, wherein
an attacker tampers with air traffic data to cause a TCAS induced
collision \cite{ref52}.

\subsection{TCAS Attack }

Despite TCAS strong reputation regarding safety, it was not designed
to withstand any type of attacks, i.e. jamming \cite{ref34}. One
of the general ways is to jam 1090 MHz channel to prevent the aircraft
from tracking potential “intruders”. However, these types of attacks
are easily detected and countered \cite{ref34,ref36}. More effective
jamming attacks include so called “All-Call Flood” and “Squitter Flood”.
During the “All-Call Flood” the attacker takes advantage of All-call
interrogation and occupies 1030 MHz channel to trigger all nearby
Mode S transponders to reply with their 24-bit International Civil
Aviation Organization (ICAO) address and flood the 1090 MHz reply
channel. The “Squitter Flood” attack is performed by an attacker spoofing
the nearby transponders by transmitting replies on 1090 MHz channel
and forcing them to continuously track the “false” aircraft. Both
of these attacks increase the chances of Near Midair Collision (NMAC)
events however the attacker does not have full control over the NMAC
occurrence. Different kind of attack that is not related to channel
flooding is called “Phantom Aircraft” attack. If the attacker can
produce an accurate Mode S reply and seem to move like an airplane,
the TCAS transponder assumes that these replies are coming from an
actual aircraft, which will force it into tracking. Such false tracking
can lead to generation of RAs that could lead to an NMAC \cite{ref34}.

\subsection{UAV vs Manned Aircrafts}

Traditional aircraft rely on secure and redundant avionics systems,
including protected ATC channels and hardened cockpit systems, which
reduce their susceptibility to cyberattacks \cite{ref03,ref19,ref42}.
In contrast, UAVs depend on unmanned control links, such as SATCOM,
Wi-Fi, 4G/5G, or RF signals, making them highly vulnerable to jamming,
signal spoofing, and interception. A compromised control link in UAVs
can result in hijacking (e.g., GPS spoofing) or complete mission failure
\cite{fonseca2021identifying,humphreys2013detection}. Manned aircraft
incorporate hardened avionics systems, including shielded onboard
computers, encrypted flight control software, and physically secured
cockpits, which mitigate the impact of cyber intrusions. In contrast,
UAVs particularly commercial and consumer-grade drones often lack
secure hardware architectures and may rely on open-source or commercially
available flight controllers \cite{kong2024uav,ref85}. This lack
of security makes them more susceptible to firmware exploitation,
malware injection, and unauthorized software modifications. UAVs are
inherently more vulnerable than manned aircraft due to several factors,
including the absence of human intervention, heavy reliance on wireless
communication, and the presence of unhardened systems in commercial
drones, which often lack robust cybersecurity measures. Furthermore,
autonomous decision-making introduces additional risks, such as adversarial
AI attacks that could disrupt UAV navigation and mission planning,
as well as vulnerabilities in swarm coordination that may compromise
entire fleets. As UAV adoption increases, enhancing their cyber resilience
is essential for ensuring mission security in both civilian and military
applications. This can be achieved through the implementation of secure
AI, encrypted communication protocols, anti-jamming mechanisms, and
blockchain-based authentication systems.

\section{UAV-to-Command Center Coordination Attacks\label{sec:UAV-to-Command}}

\begin{figure*}[h]
	\centering{}\includegraphics[scale=0.56]{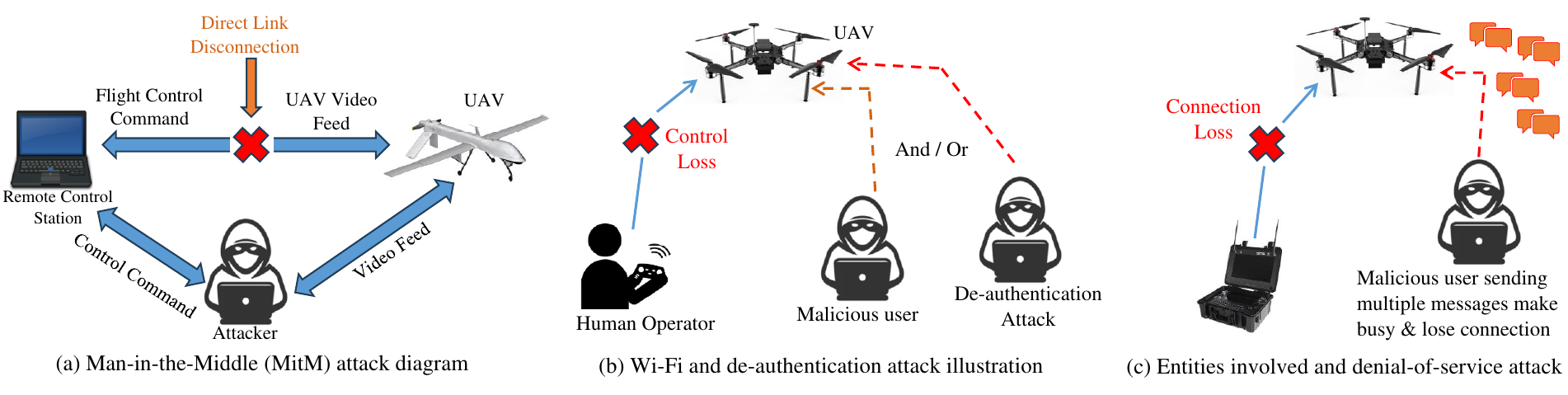}\caption{\label{fig:Malicious_attack}UAV-to-command center coordination attacks:
		(a) man-in-the-middle attack; (b) Wi-Fi attack and/or de-authentication
		attack, wherein a malicious attacker closes the communication channel
		used by a legitimate user; and (c) Entities involved in a DoS attack.}
\end{figure*}

\paragraph*{Eavesdropping Attack}It is a passive yet highly insidious
form of cyber threat in UAV networks. In this type of attack, malicious
actors clandestinely listen to and record unencrypted data transmissions
between UAVs and their ground control centers. While this attack does
not directly disrupt the operation or functionality of the UAV network,
its implications can be severe. By intercepting sensitive information,
such as navigation commands, system telemetry, or operational parameters,
attackers can gain critical insights into the network's architecture
and operations. This stolen information can then serve as a foundation
for more sophisticated and active attacks, such as spoofing, jamming,
or hijacking the UAV system. The passive nature of eavesdropping makes
it particularly challenging to detect, emphasizing the importance
of implementing robust encryption protocols and secure communication
channels to safeguard against such vulnerabilities \cite{ref55}.

\paragraph*{Man-in-the-Middle Attack}The approach man-in-the-middle
attack provides an attacker with complete control over the communication
channel between a UAV and its command center (illustrated in Figure
\ref{fig:Malicious_attack}.(a)). Passively, this allows attackers
to eavesdrop on information sent from the UAV to the command center
and vice versa \cite{ref61}. Since the attacker also has control
over the communication link, they can masquerade as an actual user
and communicate with the UAV or command center, potentially appearing
as a threat depending on the attacker’s intention \cite{ref50}. 

\paragraph*{Jamming Attacks}In jamming attacks they typically target
the ability of the communication channel between a UAV and ground
control station, effectively rendering the two entities to appear
disconnected. In most cases, the UAV’s protocols for loss of link
would become active, in which the UAV emits messages searching for
the ground control station. As long as the jamming attack persists,
the UAV is unable to re-establish communication with the ground station.
This prevents the UAV from executing its intended mission, allowing
attackers to nefariously exploit the situation \cite{ref51}. All
wireless networks are vulnerable to jamming attacks \cite{ref62}.
This issue stems from the fact that commercial software-defined radios
can be easily programmed to function as a jammer, whose effects must
be dealt with at the physical network layer. Jamming attacks can affect
not only transponder signals, but also any communication network that
the UAV relies on such as GPS, satellite communications (SATCOM),
or even cellular networks \cite{ref62}. Jamming attacks are typically
classified by how the jamming is performed, while more specific classifications
also incorporate the network protocol being used. A generic classification
of jamming attacks is presented by \cite{ref62}: 
\begin{itemize}
	\item Constant Jamming Attacks -- An attacker broadcasts a powerful jamming
	signal at all times. This continuously occupies the channel between
	the transmitter and receiver, preventing any information exchange.
	The target frequency can be adjusted depending on the attacking process,
	occupying either the entire channel bandwidth or only a fraction of
	it. 
	\item Reactive Jamming Attacks -- An attacker broadcasts a jamming signal
	only when it detects legitimate packets being transmitted in the channel.
	This is more energy efficient than a constant jamming attack but requires
	higher performance equipment to be able to react and send a jamming
	signal in response to a packet.
	\item Deceptive Jamming Attacks -- An attacker sends meaningful signals
	to the receiver in order to waste the receiver’s time and resources,
	preventing real users from accessing the channel. This is most often
	performed for Wi-Fi networks. 
	\item Random and Periodic Jamming Attacks -- An attacker sends jamming
	signals for random periods of time and then remains idle for the rest
	of the time. This is more energy efficient than constant jamming but
	is less effective at disrupting real transmissions on the channel. 
	\item Frequency Sweeping Jamming Attacks -- An attacker switches between
	jamming signals at different channel frequencies. This is a workaround
	used by lower-cost jammers that cannot attack many channels simultaneously,
	effectively allowing for lower quality hardware to interfere with
	more channels. 
\end{itemize}
These generic jamming techniques are applicable to all UAV networks,
but the most frequented cases are around GPS jamming. Due to the weak
signal strength that GPS signals are received, jamming these signals
can be very easily performed by broadcasting noise near the GPS frequency
band. A short-range experiment was conducted in \cite{ref52,ref62},
in which an \$8 USD GPS jammer which was purchased from eBay was used
to demonstrate the ease and accessibility of GPS jamming hardware. 

\paragraph*{Wi-Fi Attack}A more recent cyberattack against UAVs which
makes use of Wi-Fi signals to disrupt the communication channel between
a UAV and its ground station, and potentially commandeer the aircraft
\cite{ref64}. Only UAVs that operate on a Wi-Fi signal are susceptible
to this type of attack, however given the rising popularity of Wi-Fi
for use on drones, this lends itself to becoming a more viable attack
strategy (demonstrated in Figure \ref{fig:Malicious_attack}.(b)).
The attacker first selects a target Wi-Fi network and attempts to
gain authorization by launching a de-authentication attack \cite{ref65}.
Once the bypass is successful, the attack is able to hijack the UAV
over the Wi-Fi network \cite{ref51,ref50}.

\paragraph*{Large-scale UAV networks attacks}Large-scale UAV networks
rely on interconnected communication protocols, including ad hoc networks
(UAV-MANETs), mesh networks, and cloud-based control systems \cite{ref77}.
While these architectures enhance operational efficiency, they also
introduce attack vectors that can compromise the entire network. A
single compromised node can escalate into a widespread security breach,
enabling various cyber threats. Malware or malicious commands can
propagate through wireless communication links, potentially disrupting
the entire UAV fleet and causing a DoS attack \cite{ref70}. Exploiting
a single vulnerability in UAV firmware can lead to swarm-wide control
loss, facilitating mass hijacking. Additionally, adversaries can intercept
and modify command signals, resulting in UAV deviations, mission failures,
or complete system takeovers. Furthermore, attacks on the swarm coordination
system's consensus mechanism may lead to swarm desynchronization,
formation collapse, or collision events, significantly impacting mission
success and operational safety.

\paragraph*{False Positive and False Negative Rates}The false positive
rate (FPR) and false negative rate (FNR) are critical metrics in evaluating
the effectiveness of intrusion detection systems (IDS) for various
cyberattacks. In Eavesdropping Attacks, where an attacker intercepts
confidential data transmissions, IDS may generate false positives
due to legitimate users engaging in high-volume data transfers, while
false negatives can occur if the eavesdropping is conducted using
passive techniques that do not alter network traffic patterns \cite{ref55}.
Similarly, man-in-the-middle Attacks involve intercepting and modifying
communication between two parties. IDS can mistakenly flag legitimate
proxy-based communications as attacks (false positives), while sophisticated
man-in-the-middle techniques using encryption tunneling may go undetected
(false negatives) \cite{ref50}. For Jamming Attacks, which aim to
disrupt wireless communication by overwhelming channels with noise,
the false positive rate can rise due to environmental interference
(e.g., microwave signals or overlapping Wi-Fi channels), whereas a
high false negative rate may result from adaptive jamming techniques
that blend into normal traffic patterns \cite{ref62}. Wi-Fi Attacks,
such as deauthentication or rogue access points, can also suffer from
high FPR if network administrators frequently reconfigure access points,
while FNR can be high when attackers use Media Access Control (MAC)
address spoofing to disguise their presence. In Large-scale UAV Network
Attacks, including GPS spoofing and signal hijacking, the false positive
rate may increase if benign anomalies (e.g., sudden wind changes affecting
UAV stability) are misclassified as attacks, while a high false negative
rate may arise due to attackers leveraging advanced machine learning-based
evasion techniques to bypass detection systems \cite{ref70}.

\section{UAV Functionality Attacks\label{sec:UAV-Functionality-Attacks}}

\paragraph*{UAV Cybersecurity Standards}To ensure the safe and secure
integration of UAVs into airspace and optimal UAV functionality, regulatory
bodies such as the FAA in the U.S. and the EASA have established cybersecurity
guidelines \cite{naik2020dilemma,wanner2024uav}. The FAA's Special
Condition for small UAVs with Remote Identification mandates that
UAVs transmit unique identifiers to enhance accountability and prevent
unauthorized drone activities. Additionally, the FAA’s UAV Traffic
Management framework incorporates cybersecurity requirements to protect
UAV communication channels from spoofing, jamming, and unauthorized
access \cite{naik2020dilemma,wanner2024uav}. Similarly, EASA's Regulation
outline cybersecurity and data protection measures for UAV operations,
requiring encryption, secure authentication, and compliance with ISO/IEC
27001 cybersecurity standards to safeguard drone networks from cyber
threats \cite{shafique2021survey,pirayesh2022jamming}. In addition
to regulatory frameworks, the Institute of Electrical and Electronics
Engineers (IEEE) has developed protocols to standardize UAV cybersecurity
practices. The IEEE 802.11s protocol enhances secure UAV-to-UAV communication
in mesh networks, reducing vulnerabilities to eavesdropping and interference.
Meanwhile, IEEE 1609.2 defines security mechanisms for vehicular and
UAV-based communication, ensuring encrypted and authenticated message
exchanges in safety-critical applications \cite{ref66,ref70}. Furthermore,
the IEEE P1920.2 Standard for UAV Swarm Communication and Security
provides guidelines for mitigating cyber risks in large-scale UAV
deployments, incorporating cryptographic techniques and blockchain-based
identity management \cite{hawashin2024blockchain,chakraborty2024blocktoll}.
Aligning with these standards ensures that UAV cybersecurity frameworks
remain resilient, interoperable, and compliant with industry best
practices.

\paragraph*{Exploitation of Recorded Video}The exploitation of recorded
video is a general category pertaining to an attack on the subset
of UAVs that use a camera to navigate and avoid collisions. This type
of attack firstly requires the attacker to have backdoor access to
the UAV’s flight controller, which can be achieved through trojans,
malware, or other means. This allows the attacker to access the flight
controller and systematically replace the real-time camera footage
of the drone with a fake view, leading the drone to believe the false
video is in fact real. In many cases, the intention of this attack
is to trick the drone to land in a different location, allowing attackers
to steal the drone \cite{ref51,ref50}.

\paragraph*{Denial-of-Service}A DoS attack is a standard term used
to describe intentional communication attacks on a receiver, often
by overloading the receiver with information that then forces the
system to halt (depicted in Figure \ref{fig:Malicious_attack}.(c)).
However, in general, denial of service can also include attacks that
force the hardware system to work overtime, leading to issues such
as memory and Central Processing Unit (CPU) congestion, buffer overflows,
and battery exhaustion \cite{ref50,falahati2022improve}. Malware,
such as hardware trojans or those previously discussed, can lead to
denial of service attacks that cause the UAV to malfunction. Such
problems, when done at the right time, can lead to the UAV landing,
crashing, and shutting down at unsafe times \cite{ref51}. In the
context of UAV Wi-Fi networks, examples of DoS attacks include Transmission
Control Protocol/User Datagram Protocol (TCP/UDP) flooding, as well
as de-authentication attacks \cite{ref65}. A TCP/UDP flood involves
sending a barrage of TCP or UDP packets to the UAV’s corresponding
network port, effectively overloading the UAV’s processing capability
and rendering the UAV unable to respond or perform its tasks. This
manner of DoS attack is similar to that applied in regular computer
networks \cite{ref66}. A de-authentication attack involves an attacker
impersonating a real user transmitting a de-authentication request,
which results in the UAV ceasing communication with the real user
and crashing due to loss of controls. The de-authentication request
is standardized in the IEEE 802.11 protocol, which is used for Wi-Fi
networks, and is meant to allow access points such as the UAV to save
on computational resources when a user wishes to end the communication
link. However, when this is triggered by an attacker, the user loses
control over the UAV, potentially leading to disastrous outcomes \cite{ref65}.

\begin{figure}[h]
	\centering{}\includegraphics[scale=0.6]{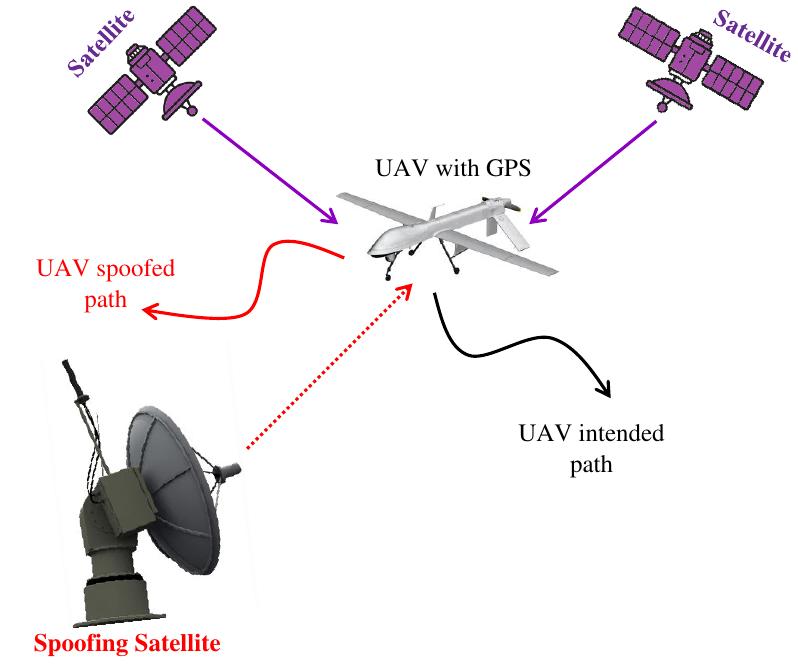}\caption{\label{fig:Spoofing2}GPS spoofing attack.}
\end{figure}

\paragraph*{GPS Spoofing}Similar to the other cyberattacks related
to message injection, GPS spoofing involves creating a fake GPS signal
which is broadcasted at UAVs, causing them to acknowledge a faked
position (see Figure \ref{fig:Spoofing2}). The artificial GPS signals
can be created by either ground equipment, or actual satellite that’s
broadcast at a higher power than GPS satellites. Once the drone is
given its falsified position, it can then be hijacked, leading to
theft or crashes \cite{ref51}. Various types of spoofing techniques
exist, although the general methodology remains consistent \cite{ref67}:
\begin{itemize}
	\item Simple Spoofing -- Generating fake Global Navigation Satellite Systems
	(GNSS) signals for transmission. Can be implemented using low-cost
	hardware to receive and reproduce GNSS signals, or commercial hardware
	with greater processing capability. Typically, GNSS is very sensitive
	to spoofing attacks because of weakness of satellite signals at the
	earth’s surface, in particular these signals are used publicly and
	not protected.
	\item Spoofing with high gain antennas -- The attacker uses high gain antennas
	to separate the GNSS signals from noise.
	\item Intermediate Spoofing -- It is an attack through a receiver-spoofer
	and the attacker generates false GNSS signals while simultaneously
	attempting to attack the target receiver through code phase alignment
	between the fake and real signals. The receiver will track the satellite
	signals to synchronize as accurate as possible using the satellite
	time and estimate the Doppler frequencies and code phases of each
	satellite signal tracked by the victim receiver.
	\item Spoofing with multiple transmitting antennas -- An advanced technique
	used against receivers that have multiple antennas. The attacker uses
	multiple antennas to attack each of the receiver’s antennas. Every
	transmitting antenna of the attacker is directly related to corresponding
	receiving antenna on the victim side.
	\item Sophisticated spoofing -- Performed by groups of attackers, who coordinate
	and synchronize their attacks on the receiver GPS system. They are
	able to attack the victim's receiver in an efficient as well as organized
	manner. They can have 3D position data of the victim's antenna and
	in turn can overcome complex countermeasures, for instance angle of
	arrival estimation.
\end{itemize}
This is much more effective at overcoming complex anti-spoofing countermeasures,
such as techniques that employ estimation of angle of arrival. The
process of creating a portable GPS spoofing system with low-cost SDR
equipment and publicly available code frameworks is presented in \cite{ref67,ref68}.
The investigation concluded that, due to the lack of protection for
GPS receivers against spoofing, it was possible to spoof a receiver’s
positions quite easily. Note however that this is not legal, and should
not be attempted without authorization.

\section{UAV Electronic Warfare for Military and Civilian Purposes\label{sec:UAV-Military}}

In the above sections the discussion has focused on the civilian side
of UAV cyberwarfare, however it should be noted that those same tactics
are applicable to military UAVs as well. This section starts first
by discussing support measures, countermeasures, and counter-countermeasures.
Next, some notable UAV cyberattacks occurrences are presented.

\subsection{Support, Countermeasures, and Counter-countermeasures}

For military drones, UAV electronic warfare includes not only the
previously listed cyberattacks, but also intelligence and counterintelligence
operations employing the use of UAVs to carry out a specific signaling
mission. Intelligence related missions typically involve setting up
UAVs to spy on enemies for indefinite periods of time, acquiring information
regarding enemy movements, routines, and patterns such that future
attack can be planned with a high degree of situational awareness.
Part of this intelligence acquisition relates to safeguarding the
electromagnetic spectrum for friendly use, and disallowing enemies
from having control over the spectrum \cite{ref52}. Details of UAV
usage in the military for electronic warfare are typically not published
publicly for obvious reasons. However, given that the advantages of
using an unmanned aerial vehicle for militaristic are plentiful, it
should come as no surprise that its applications in this realm have
not been fully explored. Classical electronic warfare was divided
into categories of EM Support Measures, EM Countermeasures, and EM
Counter-countermeasures \cite{ref45}. The USA and NATO have updated
these categories to reflect modern advancements in electronic warfare
\cite{ref69}: 
\begin{itemize}
	\item ES -- Electronic Warfare Support (Formerly EM Support Measures):
	It deals with collecting enemy signals, communications or otherwise,
	for immediate action (jamming, location determination, etc). Its main
	objective is to intercept, locate, identify and/or localize source
	of unintentional radiated EMS for the sake of immediate threat.
	\item EA -- Electronic Attack (Formerly EM Countermeasures): EA also includes
	other weaponry such as anti-radiation weapons and directed-energy
	weapons. EA main goal is to attack personnel, equipment, or facilities,
	with the focus on degrading, neutralising, or destroying the enemy
	capability.
	\item EP -- Electronic Protection (Formerly EM Counter-countermeasures):
	Alongside these categories is Signal Intelligence (SIGINT) which involves
	the reception of enemy EM transmissions. SIGINT is composed of Communications
	Intelligence (COMINT -- receiving enemy communications signals to
	extract intelligence) and Electronic Intelligence (ELINT -- using
	enemy’s communications patterns to develop countermeasures) \cite{ref45}. 
\end{itemize}

\subsection{Notable Occurrences of UAV Cyberattacks }

Let us explore some of notable occurrences of UAV cyberattacks over
the last three decades. These notable instances of UAV cyberattacks
were reported from civilian and military sectors and they are listed
below:
\begin{itemize}
	\item 2013 -- Hak5, a popular cyber and informational security company,
	demonstrated numerous vulnerabilities in UAV systems, ranging from
	using drones as Wi-Fi eavesdroppers to a DoS attack that forced the
	drone to drop out of the sky. Among these experiments was one which
	allowed drones infected by malware to seek out and infect other drones
	as well, providing total takeover of the control system \cite{ref52}.
	\item 2012 -- A GPS jamming attack caused an S-100 Camcopter UAV to crash
	into the ground control van, killing a one engineer and injuring two
	remote pilots during testing. This occurred in Incheon, South Korea,
	from an unknown attacker \cite{srirangam2023safety}.
	\item 2011 -- the U.S. military’s RQ-170 Sentinel UAV was hijacked in Iran,
	leading to the drone’s capture and breach of informational security.
	Although the method of hijacking was never confirmed, researchers
	afterwards proved that it was possible to hijack drones through GPS
	spoofing \cite{vanitha2023cyber}. It is suspected that a combination
	of GPS jamming and spoofing led to the UAV being captured.
	\item 2011 -- A Nevada ground control station manning a UAV network was
	infected by a keylogger. No known classified information was lost
	or transmitted \cite{malaviya2020digitising}.
	\item 2009 -- Iraqi forces intercept a video stream being transmitted from
	a UAV to its ground control station. The unsecured communication link
	was intercepted using SkyGrabber, an off-the-shelf product for use
	with satellite feeds \cite{ref61}.
	\item 1996 -- An Israeli UAV’s video feed was allegedly intercepted by
	Hezbollah militants, assisting the forces in ambushing and killing
	Israeli commandos \cite{taneski2020use}.
\end{itemize}
In a summary of detailing instances of UAV cyberattacks, it was found
that GPS spoofing attacks were the most common method, with GPS jamming
coming in second. Less common attacks included de-authentication attacks,
zero-day vulnerabilities, exploitation of recorded video, interception
of data feeds, and virus attacks.

\subsection{Electronic Warfare impact on Civilian UAVs}

Civilian UAV applications, such as disaster response, environmental
monitoring, and precision agriculture, have witnessed significant
growth in recent years due to advancements in UAV technology and artificial
intelligence. Ensuring cybersecurity in these applications is crucial
to maintaining operational safety, protecting sensitive data, and
ensuring sustainable deployment. UAVs play a critical role in disaster
response by providing real-time surveillance, assessing damage, and
locating survivors. However, a lack of robust cybersecurity measures
can lead to system intrusions, resulting in misinformation or loss
of control. Cyberattacks on UAVs used in disaster zones can compromise
mission-critical data and endanger lives. Implementing strong encryption
protocols and secure communication channels ensures data integrity
and reliable decision-making. Research emphasizes the role of cybersecurity
in UAV systems to safeguard data transmission and maintain operational
trustworthiness \cite{wanner2024uav}. Environmental Monitoring UAVs
are increasingly used for environmental monitoring tasks, such as
tracking deforestation, wildlife, and pollution levels. These systems
collect sensitive data that, if intercepted, could be misused. Cybersecurity
frameworks protect this data, ensuring confidentiality and preventing
unauthorized access. Advances in cryptographic techniques and secure
data storage methods can mitigate the risk of data breaches. Studies
show that UAV data security is essential for maintaining the credibility
of environmental monitoring programs \cite{ref43}. In precision agriculture,
UAVs collect data on soil conditions, crop health, and resource usage,
enabling data-driven farming decisions. Cyberattacks on UAV networks
could lead to altered data or service disruptions, potentially causing
significant economic losses. Implementing cybersecurity measures,
such as multi-factor authentication and intrusion detection systems,
enhances the reliability and safety of agricultural UAV operations.
Research highlights that secure UAV systems improve overall sustainability
by ensuring data accuracy and protecting against malicious interference
\cite{alahe2024cyber}. Cybersecurity Enhancing Sustainability and
Safety Overall, cybersecurity measures not only prevent unauthorized
access and data breaches but also ensure the long-term viability of
UAV operations by maintaining system integrity. This directly contributes
to operational safety, as secure UAVs are less prone to hijacking
or failure due to cyber attacks. Moreover, ensuring data integrity
and availability supports sustainable practices, particularly in precision
agriculture and environmental monitoring, where data accuracy is paramount
for decision-making.

UAVs are widely deployed for air quality monitoring, especially in
regions prone to industrial pollution or wildfires. These drones collect
real-time data on pollutants like $CO_{2}$ and particulate matter.
However, their reliance on wireless communication makes them susceptible
to electronic warfare attacks, such as GPS jamming or spoofing. An
attack compromising their navigation systems could not only disrupt
data accuracy but also increase energy consumption through re-routing
or signal loss recovery processes. During the 2020 wildfires in California,
UAVs were instrumental in mapping air quality and guiding firefighting
strategies. An incident reported in 2017 demonstrated how GPS interference
during a test flight in Nevada affected UAV stability and fuel efficiency
\cite{guo2019covert}. This shows the dual need to secure these systems
against electronic threats to ensure reliable environmental data collection
and minimize unintended environmental impacts due to system failures.
Electronic warfare and cybersecurity operations often involve extensive
use of data centers, which have substantial energy demands. Efficient
cybersecurity practices, such as optimized encryption protocols, can
reduce the computational burden and thereby lower energy consumption.
For instance, adopting lightweight cryptographic methods can cut power
usage by $20\%$ compared to traditional encryption \cite{roma2021energy}.

\section{Countermeasures and Defensive-Aids\label{sec:Countermeasures-and-Defensive}}

The idea behind countermeasures and defensive-aids is to protect the
data handled by UAVs in terms of confidentiality, integrity, and authenticity
as well as to guarantee the availability of service \cite{hashim2023exponentially,ref71,ref73,ref74,ref79,ref80,ref84}
in both civilian and military applications. Considering the variety
of UAV cyberattacks discussed in Section \ref{sec:UAV-to-UAV}-\ref{sec:UAV-Military},
naturally over time a wide spectrum of countermeasures to these attacks
were created to cater to vulnerabilities found on hardware, software,
and network layers. Because the same countermeasure method can work
against multiple types of attacks, these methods are discussed in
broad categories: prevention, detection, communication traffic, flight
behavior, mitigation, inertial navigation (see Figure \ref{fig:Common-UAV-cyberattack}.(c)).
Each category includes specific countermeasures which are discussed
in more detail with respect to the aforementioned attacks.

\subsection{Prevention\label{subsec:6-Prevention}}

Prevention methods focus on making it difficult to attack the UAV
in the first place. The three main subgroups of preventative methods
include “Access Control”, “Information Protection” and “Component
Selection”. 

\paragraph*{Access Control}The essence of access control is to ensure
that the UAV can be contacted only by authorized personnel or software.
This can be done with password-based node authentication schemes.
One example of realization of this method, in case if Wi-Fi is used
as a communication protocol, is to establish MAC address of a device
that is trying to access the UAV as a password. MAC is a unique hardware
identifier that is attributed to each electronic device on a network,
therefore if specific MACs are predefined for access prior to mission,
any other authenticating messages will be rejected by UAV \cite{ref72}.
This type of countermeasure can prevent GPS spoofing and Wi-Fi attacks
\cite{ref73}.

\paragraph*{Information Protection}As discussed prior, the messages
involved in UAV links can be intercepted, eliminated or infused. Several
methods can be employed to protect the confidentiality and integrity
of the messages sent and received by UAVs. One of the most widely
discussed method is usage of cryptography, particularly because most
UAVs operate over insecure data links such as GPS and Wi-Fi. There
are two main types of encryption schemes: asymmetric and symmetric.
Asymmetric schemes make use of a public key (used for encrypting the
message) and the private key (used for decrypting the message) whereas
in symmetric schemes the same key is used for encrypting and decrypting
the message \cite{ref74}. The asymmetric encryption is more computationally
demanding and can be challenging to implement on UAVs with limited
resources \cite{ref71}, hence the symmetric schemes might be preferred.
Another way to interfere with message integrity is jamming. In order
to protect the link from jamming spread spectrum techniques are utilized,
such as Frequency Hopping (FH), Frequency Hopping Spread Spectrum
(FHSS), and Direct Sequence Spread Spectrum (DSSS). Spread spectrum
techniques involve using random sequences to spread the message over
a wide band of frequencies which makes jamming ineffective since it
blocks only one specific frequency at a time. The sequence is shared
between the sender and receiver which then allows for “decoding” the
message \cite{ref75}.

\paragraph*{Component Selection}The component selection strategy
relies on designing or sourcing components that do not have exploitable
vulnerabilities. This can be achieved on software, firmware/sensors
levels, however it could be quite challenging to achieve on all levels
as it could potentially drive the cost and complexity up. Instead,
some anti-tampering technologies can be employed on board of the UAVs
to prevent entry points for potential attacks \cite{ref71}.

\subsection{Detection and Radio Signal Characteristics }

Once the prevention methods fail, it is crucial for UAV to be able
to detect and identify the type of attack that it is experiencing.
The largest subgroup in detection methods is related to “presence
of anomalies”. 

\paragraph*{Presence of Anomaly}Cyberattacks often cause abnormal
patterns, a variety that can be observed in radio signals, communication
traffic, flight behaviour or the surrounding environment. These anomalies
can help detecting an attack on a UAV of interest. Cyberattacks often
target the communication link between UAVs and their control stations.
This can cause anomalies in the radio signals. Unusual changes in
signal strength might indicate jamming or spoofing attacks. The appearance
of unexpected frequencies or distortions in the spectrum can suggest
intentional interference. Prolonged delays or missing packets in telemetry
data transmission may indicate eavesdropping or DoS attacks. The data
exchanged between a UAV and its ground station or other UAVs is a
prime target for attackers. A surge or drop in data transmission could
signify an attempted intrusion or data exfiltration. Deviation from
standard communication protocols may indicate malicious attempts to
compromise the system. Access from unauthorized or suspicious IP addresses
can suggest a potential breach.

\paragraph*{Radio Signal Characteristics}GPS spoofing causes sudden
changes in received signal such as signal strength, noise level, signal
phase-delay, etc. One of the methods to detect GPS spoofing attack
is to set a threshold value for signal strength that will determine
if the incoming signal is too strong and hence is a spoofing attempt
\cite{ref71}. This simple method is also enhanced by \cite{ref77}
through dynamic changing of threshold value using machine learning
algorithm (e.g., Support Vector Machine (SVM)) which counters attacker’s
attempt to change the transmitted signal strength and bypass the threshold.

\subsection{Communication Traffic }

Message elimination and infusion may cause anomalies in the communication
traffic. For example, the data packet delivery ratio can drop significantly
because of message elimination or, on the contrary, as a result of
message injection attack there can be a surge of incoming messages
\cite{ref71}. One of the methods to detect attacks of this type is
to have a mechanism which allows the UAVs in the network to compare
their traffic parameters with neighboring UAVs. This method is also
effective against spoofing attempts \cite{ref79,naik2020dilemma}.
Additionally, machine learning techniques are used to detect general
anomalies in network traffic by taking in parameters such as packet
size, flow duration, number of packets. Study in \cite{ref80} compares
the effectiveness of various machine learning algorithms and finds
that decision tree algorithm is the most accurate compared to other
algorithms. 

\subsection{Flight Behaviour and UAV Environment}

GPS spoofing often causes changes in UAVs trajectory or other flight
parameters as its main goal is to lead the UAV away from its intended
course. If a UAV has a very predictable and repeatable path, then
an algorithm based on flight statistics can be employed to detect
any anomalies in flight profile, attitude or thrust \cite{ref81}.
This method is quite limited as it would not perform well for UAVs
that do not operate over repeated path. In addition to changes in
flight parameters, GPS spoofing can also cause errors in calculated
position coordinates. Such errors can be detected by using UAV model
estimator in conjunction with position sensors on board of UAV \cite{ref82}.
Once the error reaches certain threshold the system issues an alarm
signal. To improve the performance a Kalman filter can be added to
the estimator as proposed by \cite{ref83} in order to better deal
with uncertainties in UAV movements. Finally, the last reference that
the UAV can use to detect positional anomalies is the surrounding
environment. One of the easier methods to detect spoofing attempts
is to simply compare the coordinates and imagery outputs from the
UAV. If there is an inconsistency between the two, it can be concluded
that the UAV might be either under spoofing or video replay attack
\cite{ref72}. Another interesting method to deal with video replay
attacks is to employ models that determine the solar shadow position
of the UAV according to UAV’s location, the sun position and the current
time \cite{ref84}. Inconsistency in expected solar shadows indicate
that the video output has been hijacked by the attacker. 

\subsection{Mitigation }

As the attacks are identified the UAV can move on to mitigation strategies
to overcome the attack. The main subgroups in the mitigation methods
are Neutralize/Avoid, Redundancy and Fail-Safe Protocols. 

\paragraph*{Neutralize/Avoid the Attacker}Neutralizing the attacker
can be particularly effective when the message spoofing or eavesdropping
attack has been identified. The UAV can launch a counter-jamming attack
which would overwhelm the attacker’s receiver and discontinue the
attack. The main limitation is potential consequential jamming of
friendly UAVs nearby \cite{ref72,ref73,fonseca2021identifying}. The
study in \cite{ref85} proposed cooperative jamming where one friendly
UAV is deployed as a jammer as close as possible to the eavesdropper.
This method, however, relies on the fact that the eavesdropper’s location
is known or easy to determine, which might not be the case. Avoiding
the attacker can be realized in different ways. It can be as simple
as escaping the adversary’s RF coverage range in case of a jamming,
eavesdropping or spoofing attacks. Another avoiding strategy is to
divert the attack. It can be done by creating a device that emits
signals that closely resemble the signal qualities of UAV but transmitted
at a higher power. The concept is known as honeypot and was proposed
in \cite{ref86}. 

\paragraph*{Provide Redundancy}Redundancy can be ensured on various
levels to mitigate attacks. For instance, in case of GPS jamming the
UAV can switch to a different GNSS constellation such as Galileo or
GLONASS if it is provided with antennas suitable for different GNSS
signals. This method is deemed quite effective as it is not easy to
jam all GNSS signals simultaneously \cite{abdalla2020uav}. Redundancy
can also be provided on sensor level to prevent dispatch system attacks.
For instance, an UAV can have both an optical and a MEMS gyroscope
to ensure the attitude input in case one of the gyroscopes is compromised. 

\paragraph*{Fail-safe Protocols}Fail-safe protocols are critical
safety mechanisms designed to protect UAVs and their payloads when
communication with the ground station is lost, or other system anomalies
occur. In case UAV has lost the connection with the ground station
and is not able to achieve recovery, it can resort to predetermined
protocols such as “Return to Homebase” or “Self-destruct”, to prevent
the enemy from capturing the UAV \cite{terkildsen2021safely}. These
protocols act as the last line of defense to prevent unauthorized
access, capture, or exploitation of the UAV and its systems.

\subsection{STRIDE: A Cybersecurity Threat Modeling Framework}

Spoofing, Tampering, Repudiation, Information Disclosure, DoS, and
Elevation of Privilege (STRIDE) is a threat modeling framework developed
by Microsoft to systematically identify and mitigate security threats
in software, hardware, and cyber-physical systems including drones
\cite{d2025cyber,alexandre2023cybersecurity,branco2025cyber}. STRIDE
is widely utilized in cybersecurity to analyze potential vulnerabilities
and design appropriate countermeasures \cite{branco2025cyber}. The
framework categorizes six primary security threats: spoofing, tampering,
repudiation, information disclosure, DoS, and elevation of privilege.
To mitigate these threats, various security measures are implemented.
Spoofing is addressed through multi-factor authentication, digital
signatures, and encryption. Tampering is countered using integrity
checks, digital signatures, and cryptographic hashing. Repudiation
is managed through secure logging, audit trails, and digital signatures.
Information disclosure is mitigated by implementing data encryption
and access control policies. DoS attacks are prevented through redundancy,
network filtering, and anti-jamming measures. Elevation of privilege
is mitigated by enforcing role-based access control, adhering to the
principle of least privilege, and patching vulnerabilities. STRIDE
provides a structured approach to identifying potential cybersecurity
threats, helps to design security countermeasures before a system
is deployed, and applicable to software security and drones cyber-physical
systems. Table \ref{tab:Stride} presents security aspect of transponder-based
cybersecurity and STRIDE framework \cite{d2025cyber,alexandre2023cybersecurity,branco2025cyber}.

\begin{table}[h]
	\caption{\label{tab:Stride}Transponder-Based Cybersecurity vs. STRIDE Framework}
	
	\begin{centering}
		\begin{tabular}{>{\raggedright}p{1.4cm}>{\raggedright}p{3.1cm}>{\raggedright}p{3.1cm}}
			\toprule 
			Security Aspect & Transponder-Based Cybersecurity & STRIDE Framework\tabularnewline
			\midrule
			\midrule 
			Focus & Authentication, identification, and UAV tracking.  & Threat modeling identifies six security threats. \tabularnewline
			\midrule 
			Threat Coverage & Addresses spoofing, tampering, and repudiation threats.  & Covers spoofing, tampering, repudiation, DoS, and others. \tabularnewline
			\midrule 
			Spoofing Mitigation & Uses encrypted identifiers and authentication to prevent UAV spoofing.  & Identifies spoofing and suggests countermeasures, e.g., multi-factor
			authentication.\tabularnewline
			\midrule 
			Tampering Resistance & Implement data integrity and cryptographic measures to prevent tampering.  & Uses cryptographic mechanisms.\tabularnewline
			\midrule 
			Repudiation Prevention & Provide verifiable identity logs for auditing, reducing the risk of
			denial by malicious actors.  & Suggests maintaining secure logging and digital signatures for non-repudiation.\tabularnewline
			\midrule 
			Information Disclosure Protection & Designed for identification rather than secrecy.  & Highlights the need for encryption and prevent data leaks.\tabularnewline
			\midrule 
			DoS Mitigation & Vulnerable to jamming attacks and requires redundancy or anti-jamming
			measures.  & DoS-resistant architectures, e.g., rate-limiting, redundancy, and
			anomaly detection. \tabularnewline
			\midrule 
			Elevation of Privilege Protection & Do not directly prevent privilege escalation attacks but enforce access
			control policies.  & Suggests strict access control and privilege enforcement. \tabularnewline
			\bottomrule
		\end{tabular}
		\par\end{centering}
\end{table}

\subsection{Full Autonomy and Inertial Navigation}

\paragraph*{Avoid Wireless Communication}UAVs are equipped with exteroceptive
and proprioceptive sensing units to facilitate trajectory planning,
navigation, and tracking control. UAVs are typically attached with
a variety of sensors, such as, GPS, IMU, laser lightning and Light
Detection and Ranging (LiDAR), vision units (stereo, monocular, or
RGB-D cameras), Ultra-wideband (UWB) tags and anchors, and ultrasonic
sensors \cite{hashim2023uwbITS,hashim2021geometricGPS,el2020inertial}.
GPS sensor enables satellite-based navigation providing a vehicle
navigating in 3D space with its position, heading/direction, speed,
and time information given the availability of at least four different
satellites (for 3D navigation) or at least three different satellites
(for 2D navigation) \cite{hashim2023uwbITS,hashim2021geometricGPS}.
However, GPS tends to be unreliable in certain scenarios, for instance,
GPS is unavailable indoor and it is subject to obstructions, multipath,
fading, and signal-denial \cite{hashim2023uwbITS,hashim2021geometricGPS}.
GPS also has additional challenges such as receiver clock error, satellite
clock error, Tropospheric delay, Ionospheric delay, receiver noise,
satellite orbital (ephemeris) errors, and errors due to satellite
geometry. IMU is used for orientation determination of the vehicle.
IMU solely can be used for dead reckoning (determination of orientation,
position, and linear velocity) to avoid wireless communication, however
it is only reliable for short-path navigation, however unreliable
for long-path navigation due to error drift and accumulation \cite{hashim2021gps}.
UWB sensors have similar concept to satellite-based positioning. UWB
tag(s) can provide the UAV with its position in 3D space via predefined
fixed or moving anchors using wireless communication \cite{hashim2023uwbITS}.
The main limitation of UWB units is their susceptibility to measurement
noise \cite{hashim2023uwbITS}. Onboard UAV vision units offer position
localization without wireless communication. Vision units work by
instantaneous feature points tracking of two different frames (captured
by stereo camera) or via two consecutive frames (obtained by monocular
camera) to determine vehicle's position with respect to a known earth-frame
\cite{hashim2021geometricGPS,hashim2021gps,hashim2022landmark}. Vision
units could by challenged in the event of low-texture environments
or in case of high altitude flights which may result in degradation
in positioning accuracy. LiDAR is a distance measurement sensor which
has a similar concept to radar such that instead of employing radio
waves. LiDAR utilizes light that hits the target and reflects and
helps in distance determination. LiDAR can face challenges through
the necessity of large amount of data to provide high accuracy. Also,
LiDAR faces difficulties in unstructured areas (e.g., machinery zone
and shelving). In addition, LiDAR is an expensive unit when compared
to other navigation sensors which makes it unfit for low-cost UAVs.

\begin{table*}[h]
	\caption{\label{tab:Summary-of-cyberattack}Summary of cyberattack threat focus
		and difficulty.}
	
	\centering{}%
	\begin{tabular}{>{\raggedright}p{2.5cm}l>{\raggedright}p{5.5cm}>{\raggedright}p{7.5cm}}
		\toprule 
		Attack & Type & Threat Target & Level\tabularnewline
		\midrule
		\midrule 
		Dispatch System Attack & Active & Confidentiality, Integrity, Availability, Privacy & High -- Requires installation of malware or trojans designed for
		the UAV.\tabularnewline
		\midrule 
		ADS-B Attack & Active & Confidentiality, Integrity, Privacy & Medium -- Used commercially but requires knowledge of ADS-B transmission
		protocols.\tabularnewline
		\midrule 
		TCAS Induced Collision & Active & Integrity, Availability & High -- Requires another attack (malware or misguidance) to enable
		collision to occur.\tabularnewline
		\midrule 
		TCAS Attack & Active & Availability & High -- Attacker must be familiar with the equipment ( transmit TCAS
		messages) and protocols.\tabularnewline
		\midrule 
		Eavesdropping & Passive & Confidentiality, Privacy & Low -- Can be done using receiver such as SDR. Easier if communication
		channels are not encrypted.\tabularnewline
		\midrule 
		Man-in-the-Middle Attack & Passive & Confidentiality, Integrity, Availability, Privacy & High -- Attacker must insert a device between UAV and ground stations.\tabularnewline
		\midrule 
		Jamming & Active & Availability & Low-Med -- Can be done by transmitting high powered signals at set
		frequencies. More difficult if sophisticated methods are used.\tabularnewline
		\midrule 
		Wi-Fi Attack & Active & Confidentiality, Availability, Privacy & Med -- Requires knowledge of the Wi-Fi protocol and a transmitter.\tabularnewline
		\midrule 
		Exploitation of Recorded Video & Active & Confidentiality, Integrity, Privacy & High -- Requires the attacker to install malware on the UAV and relay
		false videos to the kernel.\tabularnewline
		\midrule 
		Denial of Service & Active & Availability & Med -- Attacker can spam UAVs with useless packets. Requires a transmitter
		and other hardware.\tabularnewline
		\midrule 
		GPS Spoofing & Active & Integrity, Availability & Low-Med -- Attacker must transmit GPS signals. Such devices are available
		in commercial markets.\tabularnewline
		\bottomrule
	\end{tabular}
\end{table*}

\paragraph*{Multi-sensor Fusion}The sensor fusion helps to fuse different
sensors to improve full autonomy and enhance navigation accuracy,
and in turn it helps to accomplish what a single sensor cannot do.
In navigation, multi-sensor fusion is coupled with estimator or filter
design to reject sensor uncertainties and provide good estimation
accuracy including the hidden states such as the vehicle's linear
velocity \cite{hashim2021gps}. Examples of multi-sensor fusion applied
in UAV avionics applications for navigation purposes include GPS-IMU
\cite{girbes2021asynchronous}, vision-based navigation (vision unit
+ IMU) \cite{hashim2021geometricGPS,hashim2021gps,hashim2023exponentially,hashim2023observer},
UWB-IMU \cite{hashim2023uwbITS}, LiDAR-based IMU \cite{zhang2021lilo},
and others. As mentioned previously combination of different types
of multi-sensor fusion are used to improve estimation accuracy and
reliability. Note that each sensor has its own frequency of data measurement
collection (typical low-cost IMU has a rate of 200 Hz and low-cost
stereo camera can provide photographs at a rate of 20 Hz) \cite{hashim2021geometricGPS,hashim2021gps}.
Thus, multi-sensor fusion and filter design should account for data
collection variation frequency of different sensors and signal processing.

\subsection{Case Studies}

Cyberattacks, such as GPS spoofing or jamming, could misdirect UAVs
or disrupt communication, delaying critical assistance. A case study
demonstrated that implementing encryption protocols and anti-jamming
technologies significantly enhances UAV resilience during disaster
management missions, ensuring continuous operation and accurate data
transmission \cite{pirayesh2022jamming}. In terms of logistics and
delivery services, several companies are testing UAV-based delivery
services, where secure data transmission is crucial for package tracking
and delivery confirmation \cite{betti2024uav}. Man-in-the-middle
(MITM) attacks could intercept or manipulate this data, leading to
package theft or misdelivery. Robust encryption and authentication
mechanisms are vital to counter such threats. UAV communication frameworks
in a case study, showing how Transport Layer Security (TLS) protocols
prevent unauthorized access and data tampering in drone delivery networks,
thereby safeguarding logistical operations \cite{tu2024security}.
With regard to environmental monitoring, UAVs monitor environmental
parameters such as air quality, wildlife movements, and deforestation.
These missions depend on the integrity and accuracy of collected data.
Cyberattacks targeting data integrity could manipulate environmental
data, leading to incorrect assessments or policy decisions. In a real-world
scenario, the use of blockchain technology to ensure data integrity
in environmental monitoring UAV networks \cite{kumar2021blockchain}.
Blockchain's decentralized nature helps prevent unauthorized data
alterations, ensuring that collected environmental information remains
trustworthy and verifiable.

\section{Comparison of UAV Cyberattacks and Countermeasures\label{sec:Comparison-of-UAV}}

In this section, the cyberattacks and their countermeasures are compared
in terms of their effectiveness and ease of implementation on UAVs.

\subsection{Comparison Between Different Cyberattacking Strategies}

The various cyberattacking strategies presented in Section \ref{sec:UAV-to-UAV}-\ref{sec:UAV-Military}
were detailed in a high-level overview of the cyberattacking methodologies
most applicable to UAVs. This section elaborates more on those cyberattacks
by providing a comparison of their attacking strategies, threat focus,
and difficulty of execution. It should be noted that these cyberattacks
are not specific towards UAV or transponders, however the discussion
will still revolve around the usage of attacks on UAVs. Table \ref{tab:Summary-of-cyberattack}
shows a brief comparison of the various cyberattacks that were mentioned
in previous sections \cite{kim2020drone,shafique2021survey,mitchell2013adaptive,mpitziopoulos2009survey,he2017drone,semal2018certificateless,javaid2012cyber,chauhan2016detail,ossamah2020blockchain}.
These attacks comprise the main category of attacks, meaning more
granular subcategories would be able to identify more attacks than
those listed here.

In the context of UAVs, most cyberattacks focus on actively violating
the confidentiality of the UAV or impairing its availability. In most
cases, this is done through wireless means such as through radio waves,
such as for ADS-B attacks, TCAS induced collisions, jamming, Wi-Fi
attacks, and denial of service. These wireless attacks are more prevalent
since modern SDR are capable of transmitting or receiving at a wide
range of frequencies and can be programmed to transmit complex protocols
by malicious users. For instance, some SDRs \cite{ref87} available
online even contain pre-programmed transmission and receive protocols
for ADS-B, ACARS, etc. Such devices are typically not barred from
country imports due to the very general use of SDRs, however malicious
users are capable of using these devices for nefarious purposes.

\begin{table}[h]
	\caption{\label{tab:Prevention-comparison}Prevention methods comparison.}
	
	\centering{}%
	\begin{tabular}{>{\raggedright}p{1.7cm}>{\raggedright}p{2.4cm}>{\raggedright}p{3.6cm}}
		\toprule 
		Prevention & Attack Types & Limitation\tabularnewline
		\midrule
		\midrule 
		Node Authentication & Wi-Fi attacks, GPS spoofing, Dispatch system attack & Easy to implement solution, however, effectiveness depends on password
		complexity\tabularnewline
		\midrule 
		FHSS and DSSS & Jamming & Effectiveness diminishes as number of UAVs increases\tabularnewline
		\midrule 
		Cryptographic Encryption & ADS-B Attack, Eavesdropping, Man-in-the-Middle & Computationally demanding (limited on board computational power)\tabularnewline
		\midrule 
		Tamper-proof components & Dispatch System Attack & Additional cost, difficulty sourcing completely tamper-proof devices\tabularnewline
		\bottomrule
	\end{tabular}
\end{table}

\begin{table}[h]
	\caption{\label{tab:Detection-and-mitigation}Detection and mitigation methods
		comparison.}
	
	\centering{}%
	\begin{tabular}{>{\raggedright}p{1.7cm}>{\raggedright}p{2.4cm}>{\raggedright}p{3.6cm}}
		\toprule 
		\multicolumn{3}{c}{Prevention comparison}\tabularnewline
		\midrule 
		Prevention method & Attacks & Limitation\tabularnewline
		\midrule
		\midrule 
		Threshold method & GPS spoofing & Easy to bypass by an adversary\tabularnewline
		\midrule 
		Machine learning & ADS-B attack,
		
		Message deletion,
		
		GPS spoofing,
		
		DoS attack & Can be computationally demanding\tabularnewline
		\midrule 
		Flight statistics & GPS spoofing & Relies on predictable UAV trajectory\tabularnewline
		\midrule 
		Reference based & GPS spoofing & Can be computationally demanding\tabularnewline
		\midrule 
		Image comparison & Message injection,
		
		GPS spoofing & High dependence on communication links (UAV- to-Ground or UAV-to-UAV)\tabularnewline
		\midrule
		\midrule 
		\multicolumn{3}{c}{Mitigation comparison}\tabularnewline
		\midrule 
		Mitigation method & Attacks & Limitation\tabularnewline
		\midrule
		\midrule 
		Jamming counterattack & Eavesdropping, message spoofing/injection & Jamming could potentially interfere with the UAV itself\tabularnewline
		\midrule 
		Escaping the coverage range of an adversary & Jamming, spoofing, message injection & Uses limited power resources\tabularnewline
		\midrule 
		Adding multiple antennas for redundancy & Jamming, sensor (dispatch system) attack & Physical limitations of UAV (available space, power etc.)\tabularnewline
		\midrule 
		Implementation of fail-safe protocol & Jamming, ADS-B attack, Dispatch system attack & Last resort means (provides no link or data recovery)\tabularnewline
		\bottomrule
	\end{tabular}
\end{table}

\subsection{Comparison Between Countermeasure Strategies }

This subsection provides a comparison summary of all countermeasure
strategies discussed in Section \ref{sec:UAV-to-UAV} in terms of
attacks that they are effective against as well as their limitations.
Table \ref{tab:Prevention-comparison} provides summary of all preventive
countermeasures, where it can be seen that the most versatile preventative
methods are node authentication and encryption. In context of UAVs,
the only concern is the limited computational power which can be a
significant limitation for employing cryptographic methods on board
of smaller UAVs. The upper part of Table \ref{tab:Detection-and-mitigation}
presents summary of detection methods, which shows that machine learning
algorithms proved to be quite effective for detecting multiple types
of attacks by using RF signal and communication traffic data. Just
as in case with encryption the only limitation for implementation
of machine learning algorithms on UAV is their limited on-board computational
power. Finally, the lower part Table \ref{tab:Detection-and-mitigation}
provides with comparison of mitigation methods. All of these methods
are effective against multiple types of attacks; therefore, it is
not possible to provide a simple recommendation on which mitigation
method is best. 

Node authentication and encryption are highly versatile, as they address
a wide range of cyber threats, including eavesdropping, spoofing,
and data tampering. Smaller UAVs often lack the processing capacity
to execute complex cryptographic protocols in real-time. Implementing
robust encryption may introduce delays in time-sensitive applications
such as navigation or real-time surveillance. Machine learning-based
methods stand out for their ability to detect a wide array of attack
vectors, including jamming, spoofing, and unauthorized access. Training
and deploying machine learning models require significant computational
resources, which are often unavailable on smaller UAVs. Machine learning
systems can misidentify benign anomalies as threats, potentially disrupting
operations unnecessarily. All mitigation methods are effective against
multiple attack types; however, their suitability depends on specific
mission requirements and threat models. Advanced reactive systems
may demand significant computational and energy resources. Fail-safe
protocols, while effective, may disrupt mission objectives or result
in UAV loss under certain conditions. In general, despite providing
suggestions on more effective countermeasures, the final choice of
methods largely depends on the UAV mission and the attacks that can
pertain to that mission. For instance, in case of stationary UAV swarms
that are used for communication networks, perhaps securing the traffic
with machine learning techniques makes a lot more sense than for a
singular drone that performs delivery and would benefit more from
GNSS connectivity redundancy to ensure reliable navigation.

\section{Future Trends\label{sec:Future-Trends}}

As drones become more prevalent in our daily lives so is the risk
and complexity of cyberattacks. The current trends in the cybersecurity
have been discussed in previous sections, however as it was seen even
the most novel methods run into significant limitations associated
with limited size or resources available to UAVs. Hence this section
briefly introduces the potential future research trends that pertain
to the UAV cybersecurity.

\paragraph*{Computationally efficient machine learning algorithms}As
was shown in Table \ref{tab:Detection-and-mitigation}, the limited
computational resources available to UAVs are a significant limitation
when it comes to employing computationally heavy algorithms (i.e.,
machine learning or artificial intelligence). This can be remedied
by outsourcing training phase to the ground which would diminish the
algorithm to a simple mapping function \cite{hashim2024AdaptiveNeural}.
Such approach is limited since no new learning can be introduced online,
therefore developing computationally efficient machine learning algorithms
that can be implemented on board of a drone is an important research
vector for UAV cybersecurity.

\paragraph*{Lack of datasets for federated learning models}While
traditional machine learning methods use a centralized data source
to train the model, Federated Learning (FL) method uses multiple entities
that train a model. FL methods are actively researched in context
of UAVs since they allow for stronger data privacy protection as well
as network scalability (increase in UAV numbers) \cite{naik2020dilemma}. The biggest issue
that prevents FL models from being immediately implemented on UAVs
is the lack of drone datasets that can be used for training these
models, i.e., network traffic datasets, and malware datasets \cite{hadi2024real}.
Therefore, the future research will very likely focus on generating
more datasets that can be used for effective model training \cite{jonnalagadda2024segnet}.

\paragraph*{AI and Adversarial Machine Learning in UAV Security}AI-generated
cyber threats are becoming a significant challenge in UAV security
\cite{naik2020dilemma,li2025adaptive,li2024secure}. Attackers are
leveraging AI to automate and enhance cyberattacks, making them more
adaptive, stealthy, and difficult to detect. Adversarial Machine Learning
(AML) techniques are being integrated into UAV security models. AML-based
security frameworks can detect and neutralize adversarial AI threats
by training UAV-based IDS on adversarially generated attack scenarios.
For instance, defensive AI models can employ adversarial training
to improve robustness against evasion attacks, where attackers modify
malicious inputs to deceive detection algorithms. Furthermore, FL
and decentralized AI approaches enable UAVs to collaboratively update
their security models without exposing raw data, enhancing both privacy
and adaptability. By incorporating generative adversarial networks
(GANs) and explainable ML approaches for anomaly detection and reinforcement
learning-based defense mechanisms, UAV security systems can proactively
adapt to evolving AI-generated threats, ensuring resilient and autonomous
threat mitigation in dynamic environments.

\paragraph*{Feasible cryptographic methods}Cryptographic methods,
although are quite effective for preventing cyberattacks, run in the
same limitation as machine learning techniques, which is restricted
by computational power limitations. As was seen in Section \ref{subsec:6-Prevention}
out of two major encryption schemes the symmetric schemes are more
preferred to asymmetric ones since they are less computationally demanding.
However, symmetric schemes rely on reliable key distribution method
which in itself can require a lot of computational resources. The
work in \cite{kong2024uav} suggest distributing keys to UAV when
they are on the ground and using that one distributed key over the
duration of flight. A distribution like that would require a physically
secure channel which can be realized via Quantum Key Distribution
(QKD). Study in \cite{kong2024uav} describes implementation of BB84
QKD for UAV applications. Despite QKD being very safe, it is still
prone to man-in-the-middle attacks which is something to be addressed
in the future research.

\paragraph*{Securing data aggregation process}As the networks of
UAVs continue to expand, so does the volume of data exchanged between
UAVs and their respective ground stations. To optimize both the energy
consumption of UAVs and the associated communication costs, data aggregation
emerges as a critical process. This involves collecting and consolidating
data from multiple sources to minimize redundant transmissions and
maximize efficiency. However, ensuring the security of this aggregated
data is paramount, given the risk of interception or tampering during
transmission. Secure data aggregation protocols not only reduce overheads
but also safeguard the integrity and confidentiality of the information
exchanged. Although advancements are being made in this field, it
remains a vibrant area for future research, as evolving technologies
and threats necessitate continuous improvement and innovation in secure
and efficient data aggregation methods \cite{ref88}.

\paragraph*{Blockchain-based UAV authentication}Blockchain technology
has emerged as a robust solution for securing UAV networks by enabling
decentralized authentication and secure communication frameworks \cite{kumar2021blockchain}.
Traditional authentication mechanisms rely on centralized authorities,
which introduce vulnerabilities such as single points of failure and
susceptibility to cyberattacks. By leveraging blockchain, UAV networks
can achieve tamper-resistant, transparent, and trustless authentication,
where each UAV is registered as a node with cryptographic credentials
stored on an immutable ledger \cite{chakraborty2024blocktoll}. Smart
contracts can be used to automate authentication processes, ensuring
only legitimate UAVs can access network resources while preventing
identity spoofing and unauthorized control \cite{hawashin2024blockchain}.
This approach significantly enhances security in large-scale UAV swarms
and heterogeneous drone networks. In addition to authentication, blockchain
facilitates secure and resilient communication between UAVs and ground
control stations. By integrating consensus mechanisms such as proof-of-work,
proof-of-stake, or lightweight alternatives such as delegated-proof-of-stake,
UAV networks can verify message integrity and prevent data tampering
\cite{hawashin2024blockchain}. Furthermore, end-to-end encryption
and decentralized key management ensure that communication remains
confidential, mitigating risks from eavesdropping, man-in-the-middle
attacks, and jamming attempts. The distributed nature of blockchain
also enhances network resilience against DoS attacks, ensuring continued
UAV operations even in adversarial environments. By combining blockchain
with edge computing and AI-driven anomaly detection, UAV networks
can achieve real-time, adaptive security against emerging cyber threats.

\paragraph*{Cryptographic protections and flight endurance}Cryptographic
protections in UAVs enhance security by encrypting communication,
authenticating signals, and preventing cyber threats such as spoofing
and jamming. However, these protections impose computational overhead,
increasing the power consumption of onboard processors. This additional
energy demand can strain the UAV's limited battery capacity, reducing
overall flight endurance. The impact varies based on encryption algorithms,
hardware efficiency, and system optimization. Lightweight cryptographic
schemes and dedicated security hardware can help mitigate energy drain,
balancing security with operational efficiency. Thus, while cryptographic
protections are essential for secure UAV operations, they must be
carefully designed to minimize their impact on battery endurance.

\section{Conclusion\label{sec:Conclusion}}

Cyberattacks applicable to Unmanned Aerial Vehicles (UAVs) were comprehensively
reviewed and discussed, with their impacts and methodologies summarized
at a high level. The majority of cyberattacks targeting UAV networks
attack the system through wireless communications, with the more prevalent
attacks being those non-specific to any transponder. Examples of these
include jamming, spoofing, and eavesdropping. More sophisticated attacks
consisted of UAV recorded video attacks, man-in-the-middle attacks,
and Traffic Alert and Collision Avoidance System (TCAS) attacks. Countermeasures
focusing on the categories of prevention, detection, and mitigation
were discussed. Researched methods included stricter access control
to UAV systems, detection of anomalies in the UAV sensor readings,
and neutralization of attacks by countering the attack (i.e. for jamming)
or avoiding the attack (i.e. for spoofing) by leaving the targeted
area. As more UAVs are launched into a common airspace, it becomes
important to devise smart countermeasures to counteract the actions
performed by bad actors. Although the countermeasures listed in this
paper are mainly applicable for communication and transponder-based
attacks, the idea remains for the entire system. With newer technologies,
these methods incorporate advancements such as the use of efficient
machine learning algorithms and secure cryptography to create more
robust UAV communication networks. Overall, this paper provides a
comprehensive summary of the historical context of UAV-related cyberattacks,
countermeasures, and the evolving trends and future advancements in
these respective areas.
		
\section*{Acknowledgments}

This work was supported in part by the National Sciences and Engineering Research Council of Canada (NSERC), under the grants RGPIN-2022-04937.

\balance
\bibliographystyle{IEEEtran}
\bibliography{bib_EW_Avionics_UAV}
		
	\end{document}